\documentclass[a4paper,11pt]{article}
\pdfoutput=1 

\usepackage{jheppub}
\usepackage{bm}
\usepackage{slashed}
\usepackage[T1]{fontenc} 

\def\a{\alpha}

\def\CA{{\cal A}}

\def\CH{{\cal H}}
\def\CI{{\cal I}}

\def\CS{{\cal S}}

\def\CW{{\cal W}}

\def\CZ{{\cal Z}}

\def\U{\mathrm{U}}
\def\SU{\mathrm{SU}}

\def\SL{\mathrm{SL}}

\def\beq#1\eeq{\begin{align}#1\end{align}}

\newcommand\ptn{partition function }

\newcommand{\rG}{r_G}

\newcommand{\Tr}{\mathrm{Tr}}

\newcommand{\RR}{\mathbb{R}}
\newcommand{\ZZ}{\mathbb{Z}}

\renewcommand{\Re}{\mathrm{Re}}

\title{\boldmath Expanding 3d $\mathcal{N}=2$ Theories \\ around the Round Sphere}

\preprint{IPMU19-0184}

\author[a]{Dongmin Gang}
\author[b]{and Masahito Yamazaki}

\affiliation[a]{Asia Pacific Center for Theoretical Physics (APCTP), Pohang 790-784, Korea}
\affiliation[c]{Kavli Institute for the Physics and Mathematics of the Universe (Kavli IPMU), \\ University of Tokyo, Kashiwa, Chiba 277-8583, Japan}

\abstract{We study a perturbative expansion  of the squashed 3-sphere ($S^3_b$) partition function of 3d $\mathcal{N}=2$ gauge theories around the squashing parameter $b= 1$. Our proposal gives the coefficients of the perturbative expansion as a finite sum over the saddle points of the supersymmetric-localization integral in the limit $b \rightarrow 0$ (the so-called Bethe vacua), and the contribution from each Bethe vacua can be systematically computed using  saddle-point methods. Our expansion provides an efficient and practical method for computing  basic CFT data ($F,C_T,C_{JJ}$ and higher-point correlation functions of the stress-energy tensor)  of the IR superconformal field theory without performing the localization integrals. }

\begin{document} 
\maketitle

\section{Introduction and Summary}
\label{sec:intro}

\subsection{The Problem}

The three-sphere partition function \cite{Kapustin:2009kz,Jafferis:2010un,Hama:2010av,Hama:2011ea} 
is a powerful quantity to characterize three-dimensional $\mathcal{N}\ge 2$ supersymmetric
quantum field theories. This partition function is defined as 
the supersymmetric partition function on the ellipsoid \cite{Hama:2011ea}
\begin{align}
S^3_b := \left\{ b^2 |z|^2 +\frac{1}{b^2} |w|^2 = 1 \;:\; z,w \in \mathbb{C} \right\}\;,
\label{S3b}
\end{align}
where the ``squashing parameter'' $b$ parametrizes a family of transversely holomorphic foliations
on the three-sphere \cite{Closset:2013vra}. This geometry becomes the round 3-sphere when $b=1$:
\begin{align}
S^3_{b=1} = (\textrm{round 3-sphere})\;.
\end{align}

The supersymmetric partition function on the geometry \eqref{S3b}
depends on a set of real mass parameters $\vec{m}$ and the R-symmetry mixing parameters $\vec{\nu}$ of the theory (see eq.~\eqref{R-symmetry mixing}).
We will denote this partition function as $\mathcal{Z}_{b}(\vec{m}, \vec{\nu})$,
and the associated free energy by 
\begin{align}
F_b(\vec{m}, \vec{\nu})=-\Re\left[\log \mathcal{Z}_b(\vec{m}, \vec{\nu}) \right]\;.
\end{align}

The problem discussed in this paper is to compute a perturbative expansion of this partition function around the special point $b=1$, where the 
three-sphere \eqref{S3b} has the round metric:\footnote{There is no term linear in $b-1$, as expected from the symmetry $b\to b^{-1}$ of the geometry \eqref{S3b}.}
\begin{align}
F_b(\vec{m}, \vec{\nu})= F_{b=1}(\vec{m}, \vec{\nu}) + \frac{1}{2}(1-b)^2 F^{(2)}(\vec{m}, \vec{\nu}) +\frac{1}{3!}(1-b)^3  F^{(3)}(\vec{m}, \vec{\nu}) + \dots \;.
\label{F_expand}
\end{align}

Since integral expressions for the three-sphere partition function is already known in the literature \cite{Kapustin:2009kz,Jafferis:2010un,Hama:2010av,Hama:2011ea}, in principle this is a matter of expanding a known expression. 
The integral expression, however, is given as a complicated oscillatory integral, and this makes the expansion 
highly non-trivial and inefficient either analytically or numerically, especially if one wishes to go to higher orders in $b-1$.
The goal of this paper is to propose a different method for expanding around the value $b=1$,
which in particular does not involve any integral.

\subsection{Motivations}

There are two motivations for this problem. 

First, the quantities appearing in the expansion \eqref{F_expand} contains useful quantities characterizing the system,
and our method gives an efficient and practical method to compute these quantities.

At leading order the resulting free energy $F=F_{b=1}$  on the round three-sphere, when we choose $\vec{m}=0$ and 
$\vec{\nu}$ to be the IR superconformal R-charge, is known to decrease along the renormalization group (RG) flow \cite{Jafferis:2011zi,Klebanov:2011gs} (see also \cite{Casini:2012ei}).
At the next non-trivial order, the coefficient $F^{(2)}$ is identified \cite{Closset:2012vg,Closset:2012ru} with the ``central charge'' $C_T$,\footnote{There are counterexamples to the conjecture that $C_T$ decreases along the RG flow \cite{Nishioka:2013gza}.}
defined from the two-point function of the stress-energy tensor $T_{\mu\nu}$ \cite{Erdmenger:1996yc}:
\begin{align}
\begin{split}
&\langle T_{\mu\nu}(x)T_{\rho \sigma} (0)\rangle = C_T\frac{I_{\mu\nu, \rho \sigma}(x)}{|x|^6}\;,
\end{split}
\label{CT_def}
\end{align}
where
\begin{align}
\begin{split}
&I_{\mu\nu, \rho \sigma}(x):=\frac{1}2 \left( I_{\mu\nu}(x)I_{\rho \sigma}(x)+I_{\mu \rho}(x) I_{\nu \sigma}(x) \right) - \frac{\delta_{\mu\nu} \delta_{\rho \sigma}}3 \;,\\
&I_{\mu\nu}(x):=\delta_{\mu \nu} - 2\frac{x_\mu x_\nu}{x^2} \;.
\end{split}
\end{align}
The quantity $C_{T}$ is a useful input for the conformal bootstrap program, see e.g.\ \cite{Bobev:2015jxa}.
More generally, we can compute higher order terms in the expansion around $b=1$,
and extract more detailed information of the system, which is related to 
higher-point correlation functions of the stress-energy tensor (see e.g.\ \cite{Dymarsky:2017yzx} for related discussion in the context of conformal bootstrap, albeit in non-supersymmetric settings).

When the system has global symmetries, the three-sphere partition function depends on 
the corresponding real mass parameters $\vec{m}$. We can then consider the 
expansion of $F_{b=1}$ with respect to the parameters $\vec{m}$ around $\vec{m}=0$.
This gives $C_{J_{\U(1)_A} J_{\U(1)_B}}$, which are defined by the two-point function of conserved currents \cite{Erdmenger:1996yc},
\begin{align}
&\langle J_{I}^{\mu} (x)J_{J}^{\nu} (0)\rangle = C_{IJ} \frac{I_{\mu\nu}(x)}{|x|^4}\;,  
\label{CJJ_def}
\end{align}
where $J_{I}^\mu$ is the conserved current for the $I$-th Cartan generator of the flavor symmetry. 

The second motivation comes from the 3d--3d correspondence \cite{Terashima:2011qi,Dimofte:2011ju,Lee:2013ida,Cordova:2013cea},
which claims that the $S^3_b$-partition functions of a class of 3d $\mathcal{N}=2$ theories
are identified with the partition functions of the complexified Chern-Simons theory
on 3-manifolds (such as knot complements). The perturbative expansion at the value $b=1$
gives infinitely many topological invariants of the 3-manifold.
Note that our expansion is different from the expansion around $b=0$,
which has been studied in the context of the generalized volume conjectures \cite{Dimofte:2009yn,Dimofte:2012qj,Gang:2017cwq}.

\subsection{Main Result}

The goal of the present paper is to propose
a new, efficient method to 
systematically compute the higher-order expansion of the three-sphere partition function in $1-b$, 
and hence in particular to compute $F, C_T$ and $C_{IJ}$.
Our expression appears in eq.~\eqref{main result}. 
The resulting expression is written as a finite sum, as opposed to 
an integral (as one might expect from supersymmetric localization).

For the quantity $F$, our expression nicely matches the result in \cite{Closset:2017zgf}. 
We reproduce their result in a different approach.
Moreover our result extends the result to all-order expansion in $1-b$,
and hence to infinitely many quantities.

While we defer the detailed discussion to the following sections, 
the basis idea is to relate the expansion around $b=1$ to a different expansion
around $b=0$. Contrary to the case of the former expansion,
the integrand diverges in the latter expansion, and hence 
we can systematically compute the higher-order expansion by the saddle point method.

It is instructive to compare our discussion with the case of four-dimensional $\mathcal{N}=1$ supersymmetry.
In this case, $C_T$-maximization (or $\tau_{RR}$-maximization \cite{Barnes:2005bm} for the $\U(1)$ R-symmetry current, as related by $\mathcal{N}=1$ supersymmetry) can also be implemented by $a$-maximization \cite{Intriligator:2003jj}, 
which is a rather simple algebraic (in this case cubic) function of the trial R-charge. Our result is a similar in spirit,
but now in three dimensions.

\paragraph{Organization of the Paper}

The rest of this paper is organized as follows.
We first quickly review the supersymmetric partition functions in sec.~\ref{sec:review}.
The next section, sec.~\ref{sec:main}, contains our main result
concerning the expansion around $b=1$. We motivate this result 
from the factorization property of the supersymmetric partition function.
This property relates the expansion around $b=1$ to a different expansion around $b=0$, which 
can then be computed by saddle-point methods.
In section \ref{sec:examples} we apply our method to
several examples. In appendix \ref{App : QDL} we summarize some properties of the quantum dilogarithm function
needed for the main text.


\section{\texorpdfstring{Localization on $S^3_b$ and $S^2\times_q S^1$ : Review}{Localization on S3b and S2xS1q : Review}} \label{sec:review}
In this section, we review supersymmetric-localization of the squashed 3-sphere partition function and the superconformal index.  
\subsection{\texorpdfstring{Localization on $S^3_b$}{Localization on S3b}}
In this section, we  give a brief review of supersymmetric localization \cite{Kapustin:2009kz,Jafferis:2010un,Hama:2010av,Hama:2011ea}  of 3d $\mathcal{N}=2$ gauge theories  on  the squashed 3-sphere $S^3_b$. We choose notations that will be convenient for the discussion of perturbative expansions in later sections.

We consider a 3d $\mathcal{N}=2$ gauge theory $\mathcal{T}$ with a compact connected gauge group $G$. For simplicity, we further assume that $G$ is a product of a torus and a simply-connected Lie group. The theory can be coupled to background vector multiplets for the flavor symmetry group $G_F$ of rank $r_F$. Let Cartan generators of $G_F$ be $\{F_I\}_{I=1}^{r_F}$. The vacuum expectation values (VEVs) of the scalar fields in these multiplets are the real masses $m_I$,  $I=1,\dots,r_F$. For later use, it is  convenient to rescale the real mass parameters as follows
\begin{align}
U_I:=bm_I   \;. \label{Ufromm}
\end{align}
For the geometry $S^3_b$ to preserve some supercharges, we need to turn on the background gauge field coupled to a $\U(1)_R$ symmetry. The choice of R-symmetry is not unique and  can be mixed with flavor symmetries 
\begin{align}
R^{\vec{\nu}} =R^{\vec{\nu}=\vec{0}} + \sum_{I=1}^{r_F} \nu_I F_I  \;, \label{R-symmetry mixing}
\end{align}
and the partition function depends on the mixing parameters $\vec{\nu}$. Here $R^{\vec{\nu}} $ is the Cartan charge of the mixed R-symmetry $\U(1)^{\vec{\nu}}_R$. The mixing shifts the values of $\vec{m}$ inside the partition function:
\begin{align}
\CZ_b (\vec{m},\vec{\nu}) = \CZ_b \left(\vec{m}+ i \pi \left(b+\frac{1}b\right) \vec{\nu},\vec{\nu}=\vec{0} \right)\;. \label{mixed R-symmetry}
\end{align}
The mixing  exists (i.e.\ $\nu_I \neq 0$) only for $F_I$ which corresponds to $\U(1)$ Cartan generators of the non-simple part of the gauge group.
Note that $G_F$ includes symmetries that act on the matter as well as the topological Abelian symmetries. The topological symmetry is usually denoted as $\U(1)_J$, 
and its conserved charges are the monopole fluxes for the Abelian gauge symmetries.  This implies that real masses for the topological symmetries are the Fayet-Iliopoulos (FI) parameters.

The partition function is given by an integral of the form 
\begin{equation}
\mathcal{Z}_{b}(\vec m,\vec \nu)=\frac{1}{|\textrm{Weyl}(G)|} \int_{\Gamma_\RR} \frac{    d^{\rG} Z}{(2\pi \hbar)^{\frac{\rG}{2}}} \, \Upsilon_\hbar\left(\vec U + \left(i \pi +\frac{\hbar}{2}\right) \vec{\nu},\vec Z \right) \bigg{|}_{\vec{U} = b \vec{m}}\label{Zformula}
\end{equation}
over the Coulomb-branch parameters $Z_i$, $i=1,\dots,r_G$, where  $r_G:={\rm rank}(G)$. In this formula and further in this section we use the notation 
\begin{align}
\hbar:=2\pi ib^2 \label{hbar}\;.
\end{align}
The integrand $\Upsilon_\hbar$ at $\vec{\nu}=0$ in eq.~(\ref{Zformula}) contains the following factors:
\begin{itemize}
	\item An $\mathcal{N}=2$ chiral multiplet, that transforms with weights $(\beta,\gamma)\in (\rho_G, \rho_F)$ under the maximal tori of the gauge and the flavor groups, contributes\footnote{We use the so-called ``$k=-1/2$ regularization'' of the chiral multiplet path-integral. See e.g.\ \cite{Closset:2017zgf} for a discussion.} 
	\begin{equation}
	\prod_{(\beta,\gamma) \in (\rho_G, \rho_F)}\psi_\hbar(\beta \cdot Z+\gamma\cdot U) \;.
	\end{equation}
	 The special function $\psi_\hbar (X)$ is the quantum dilogarithm (Q.D.L). Its definition and properties are reviewed in appendix \ref{App : QDL}.
	\item An Abelian Chern-Simons term with a matrix of integer levels $k_{ij}$ contributes a factor $\exp\left(\frac{1}{2\hbar}\sum_{i,j}k_{ij}Z_iZ_j\right)$. In particular, a $\U(1)$ Chern-Simons term at level $k_{ii}$ gives $\exp\left(\displaystyle\frac{1}{2\hbar}k_{ii}Z_i^2\right)$\,, while a mixed $\U(1)\times \U(1)$ Chern-Simons term at level $k_{ij}$ gives $\exp\left(\displaystyle\frac{1}{\hbar}{k_{ij}Z_iZ_j}\right)$.
	\item FI parameter $\zeta$, which is the real mass parameter for the topological $\U(1)_J$ symmetry, contributes a factor $\exp\left(- \frac{1}\hbar ZU\right)|_{U = b\zeta}$.
	\item A level-$k$ Chern-Simons term for a simple factor of the gauge group $G$ contributes  $k_{ij}=k\langle\alpha_i^\vee,\alpha_j^\vee\rangle$ to the matrix of Chern-Simons levels for the maximal torus. Here $\alpha_i^\vee$ are the coroots of the simple factor and $\langle,\rangle$ is the canonically-normalized Killing form.
	\item An $\mathcal{N}=2$ vector multiplet contributes
	\begin{equation}
	\prod_{\lambda \in \Lambda^+_{\rm adj} }  4 \sinh \left( \frac{1}2 \lambda \cdot Z \right) \sinh \left(\frac{\pi i}{\hbar} \lambda  \cdot Z\right)\,,\nonumber
	\end{equation}
	where the product goes over the set $\Lambda^+_{\rm adj}$ of positive roots of the gauge group.
\end{itemize}

There are also contributions from the Chern-Simons terms that involve background gauge fields coupled to flavor symmetries, which we did not write out. 
In general, there are also Chern-Simons terms for the Levi-Civita and the R-symmetry connections. A careful treatment of these terms can be found e.g.\ in \cite{Closset:2017zgf,Closset:2018ghr}. We will not keep track of them but instead will define the partition function $\mathcal{Z}_b$ only up to an overall factor
\begin{align}
\textrm{exp} \left( {i \pi} \mathbb{Q}\left(b^2+ \frac{1}{b^2} \right)+ i \pi \mathbb{Q}\right) \;.\label{phase ambiguity}
\end{align}
The integration cycle $\Gamma_\RR$ in eq.~(\ref{Zformula}) is $\RR^{r_G}\subset\mathbb{C}^{r_G}$. More precisely, it is infinitesimally deformed in a suitable manner to make the integral convergent. 

\subsection{\texorpdfstring{Localization on $S^2\times_q S^1$}{Localization on S2xS1q}}
The superconformal index \cite{Bhattacharya:2008zy} is defined as
\begin{align}
\CI_q ({\bf m}_I, u_I;\vec{\nu}) := \Tr_{\CH(S^2;{\bf m}_I)} (-1)^{R^{\vec{\nu}}} q^{\frac{R^{\vec{\nu}}}2 +j_3} \prod_{I=1}^{r_F} u_I^{F_I}  \;.
\end{align}
Here $R^{\vec{\nu}}$ is the Cartan generator of  the R-symmetry $\U(1)^{\vec{\nu}}_R$ in eq.~\eqref{R-symmetry mixing}.  $\mathcal{H}(S^2;{\bf m}_I)$ is the Hilbert space of the radially-quantized 3d theory with background magnetic fluxes  $\{{\bf m}_I\}$ coupled to the Cartan generators of $G_F$ turned on the $S^2$. 
From supersymmetric localization,  the index is given as a sum/integral \cite{Kim:2009wb,Imamura:2011su}
\begin{equation}
\mathcal{I}_{q}(\vec {\bf m}, \vec{u};\vec \nu)=\frac{1}{|\textrm{Weyl}(G)|} \sum_{ \vec{\bf n} } \oint \prod_{i=1}^{r_G}\frac{dw_i}{(2\pi i w_i)}\Omega_q\left( {\bf n}_i, w_i, {\bf m}_I, u_I (-q^{1/2})^{\nu_I}  \right)  \;.\label{SCIformula}
\end{equation}
The integrand $\Omega_q(\vec{\bf n}, \vec{w},\vec{\bf m}, \vec{u})$ at $\vec{\nu}=0$ contains the following factors:
\begin{itemize}
	\item An $\mathcal{N}=2$ chiral multiplet that transforms with weights  $(\beta,\gamma)\in (\rho_G, \rho_F)$ under the maximal tori of the gauge and the flavor groups contributes
	\begin{equation}
	\prod_{(\beta,\gamma) \in (\rho_G, \rho_F)} \mathcal{I}_\Delta (\beta \cdot \vec{\bf n} + \gamma \cdot \vec{\bf m},e^{\beta \cdot \log w+\gamma \cdot \log u};q) \;,
	\end{equation}
	where the ``tetrahedron index'' $\mathcal{I}_\Delta$  is defined as \cite{Dimofte:2011py}
	\begin{align}
	\mathcal{I}_{\Delta}({\bf m}, u;q) := \prod_{r=0}^\infty \frac{1-q^{r-\frac{\bf m}2+1} u^{-1}}{1- q^{r- \frac{\bf m}2 } u }\;. \label{tetraheron index}
	\end{align}
	\item An Abelian Chern-Simons term with a matrix of integer levels $k_{ij}$ contributes a factor $\prod u_i^{k_{ij} {\bf m}_j}$.
	\item The $\U(1)_J$ topological symmetry associated to an Abelian gauge symmetry contributes  $u^{-\bf n} w^{-\bf m}$. Here, $(\mathbf{m},u)$ is a pair (magnetic flux, fugacity) for the $\U(1)_J$ flavor symmetry, while $(\mathbf{n},w)$ is for the Abelian gauge symmetry. 
	\item A level-$k$ Chern-Simons term for a simple factor of $G$ contributes  $k_{ij}=k\langle\alpha_i^\vee,\alpha_j^\vee\rangle$ to the matrix of Chern-Simons levels for the maximal torus. Here $\alpha_i^\vee$ are the coroots of the simple factor and $\langle,\rangle$ is the canonically-normalized Killing form.
	\item An $\mathcal{N}=2$ vector multiplet contributes 
	\begin{equation}
	\prod_{\lambda \in \Lambda^+_{\rm adj} } (q^{\lambda \cdot (\frac{\bf n}{2} + \log w)} - q^{-\lambda \cdot (\frac{\bf n}{2} + \log w)}) (q^{\lambda \cdot (\frac{\bf n}{2} - \log w)} - q^{-\lambda \cdot (\frac{\bf n}{2} - \log w)}) \,,\nonumber
	\end{equation}
	where the product goes over the set $\Lambda^+_{\rm adj}$ of positive roots of the gauge group.
\end{itemize}
We need to sum over all the magnetic fluxes $\vec{\bf n}$  satisfying the following Dirac quantization conditions
\begin{align}
\left\{ \vec{\bf n} \;: \; \beta \cdot \vec{\bf n} +\gamma \cdot \vec{\bf m} \in \mathbb{Z}\;, \lambda \cdot \vec{\bf n} \in \mathbb{Z}\;, \; \forall (\beta,\gamma)\in (\rho_G, \rho_F) \textrm{ and } \lambda \in \Lambda^{+}_{\rm adj}\right\}/\,\textrm{Weyl} (G)\;.
\end{align}

\section{\texorpdfstring{Expansion at $b\rightarrow 1$ from Expansion at $b\rightarrow 0$}{Expansion at b to 1 from Expansion at b to 0}}\label{sec:main}
In this section, as the main result of this paper, we propose the $b\rightarrow 1 $ expansion \eqref{main result} of the squashed 3-sphere partition \eqref{Zformula}. Using the proposed expansion, we can express correlation functions of the stress-energy tensor (including the stress-energy tensor central charge) of a 3d $\mathcal{N}=2$ theory as a finite sum over Bethe-vacua. Our proposal is motivated by the factorization property of  supersymmetric partition functions \cite{Pasquetti:2011fj,Beem:2012mb}.  Interestingly, we find that the convergent expansion  at $b\rightarrow 1$ can be reconstructed from  the asymptotic expansion at $b\rightarrow 0$.

\subsection{\texorpdfstring{Expansion at $b\rightarrow 0$  (Asymptotic Expansion) }{Expansion at b to 0(Asymptotic Expansion) }} \label{sec: b goes to 0}
In this section, we provide general strategy for computing the perturbative expansion $S_n^{(\alpha,\vec{\ell})}$ ($n \geq 0$)  of the localization integral \eqref{Zformula} in the $b\rightarrow 0$ limit around saddle point $\vec{Z}^{(\alpha, \vec{\ell})}$  in eq.~\eqref{saddle point}.  Here the index $\alpha$  labels Bethe-vacua \eqref{Bethe vacua} of the 3d theory while integer-valued vector $\vec{\ell}$ labels the unphysical shift ambiguity \eqref{gtW}. For a 3d theory associated to a hyperbolic 3-manifold \cite{Dimofte:2011ju,Dimofte:2013iv,Gang:2018wek}, the perturbative invariant $S_n^{(\alpha,\vec{\ell})} $  computes the   $n$-loop invariant of the three-dimensional $\SL(N,\mathbb{C})$ Chern-Simons theory around an irreducible flat connection $\mathcal{A}^{\alpha}$ (satisfying $d\mathcal{A}^\alpha + \CA^\alpha \wedge \CA^\alpha =0$) on the 3-manifold.  We refer to \cite{2007JGP,Dimofte:2009yn,Dimofte:2011gm,Andersen:2011bt,Dimofte:2012qj,Dimofte:2013iv,Dimofte:2014zga,Bae:2016jpi,Gang:2017cwq} for systematic study on the perturbative invariants in the context of the volume conjecture \cite{MR1434238,MR1828373}. 

\paragraph{Quantum Effective Twisted Superpotential} We define the quantum effective twisted superpotential \cite{Nekrasov:2009uh,Nekrasov:2009rc,Dimofte:2011jd,Beem:2012mb} in perturbative expansion in $\hbar$,
\beq
\mathcal{W}^{\vec{\ell}}_{\hbar}(\vec Z,\vec U,\vec{\nu})\xrightarrow{\,\,\,\hbar\rightarrow 0 \textrm{\;with fixed }\vec{U}= b \vec{m}\,\,\,}\sum_{n=0}^\infty\hbar^n\mathcal{W}^{\vec{\ell}}_n(\vec Z,\vec U,\vec {\nu})\,. 
\eeq
It can be obtained from the limit of the integrand $\Upsilon_\hbar$ in eq.~\eqref{Zformula} at $\hbar\rightarrow 0$,
\beq
\Upsilon_\hbar\left(\vec U + \left(i \pi +\frac{\hbar}2 \right) \vec{\nu},\vec Z \right)  \xrightarrow{\,\,\, \hbar\rightarrow 0 \textrm{\;with fixed }\vec{U}= b \vec{m} \,\,\,} \exp\left(\hbar^{-1}\mathcal{W}^{\vec{\ell}}_\hbar(\vec Z,\vec U,\vec{\nu})\right)\,.\label{limit}
\eeq
The contribution of Chern-Simons terms to $\mathcal{W}_n$ is clear. Most of them contribute only to $\mathcal{W}_0$, with the exception of the ones that involve the R-symmetry. Those can also contribute to $\mathcal{W}_1$. For example, a level $k$ term for a U(1) symmetry with parameter $Z$ gives
\beq
\mathcal{W}^{\rm CS}_\hbar=\frac{1}{2}kZ^2\,.
\eeq 
The contribution of a charged $\mathcal{N}=2$ chiral multiplet to $\mathcal{W}_\hbar$ can be read off from the expansion of the quantum dilogarithm function $\psi_\hbar(Z)$, which is reviewed in Appendix~\ref{App : QDL}. A chiral multiplet of charge one under a U(1) symmetry with parameter $Z$ gives
\beq
\mathcal{W}^{\rm chiral}_\hbar=\sum_{n=0}^\infty\hbar^n\frac{B_n}{n!}{\rm Li}_{2-n}(e^{-Z})\,,\quad n=0,1,\dots\;,
\eeq
where $B_n=(1,1/2,1/6,\dots)$ are the Bernoulli numbers. Finally, the W-bosons contribute
\beq
\mathcal{W}^{\rm W}_\hbar=2\pi i\rho\cdot Z+\hbar\sum_{\lambda\in\Lambda^+_{\rm adj}}\log\left(e^{\lambda\cdot Z/2}-e^{-\lambda\cdot Z/2}\right)\,,
\eeq
where $\rho=\frac{1}{2}\sum_{\lambda\in\Lambda^+_{\rm adj}}\lambda$ is the Weyl vector.

The quantum effective twisted superpotential $\mathcal{W}^{\vec{\ell}}_\hbar(\vec U,\vec Z)$ is defined up to an unphysical shift,
\beq
\mathcal{W}^{\vec{\ell}}_\hbar =  \mathcal{W}^{\vec{\ell} = \vec{0}}_\hbar+2\pi i (Z_i\ell_z^i+U_I\ell^I_u+2\pi i\ell_0)+i\pi\hbar \ell_1\,,\quad \ell^i_z,\,\ell^I_u,\,\ell_0,\,\ell_1\in\ZZ\,.\label{gtW}
\eeq
In particular, $\mathcal{W}_\hbar$ is not single-valued and has branch points at $2\pi i\ZZ$ both in $Z_i$ and $U_I$, and in going around these, $\mathcal{W}_\hbar$ gets shifted by terms of the form (\ref{gtW}).  Also in the equation (\ref{limit}), the result actually depends on the direction (phase factor of $\hbar$) in which one takes the limit, but only by terms of the form (\ref{gtW}).

There can also be contributions from Chern-Simons terms in background supergravity fields. These are physically meaningful, but we will not keep track of them. Instead, we will  allow an ambiguity
\beq
\mathcal{W}_\hbar\sim \mathcal{W}_\hbar+\pi^2\mathbb{Q}+i\pi\hbar\mathbb{Q}+\hbar^2\mathbb{Q}\,.\label{ambW}
\eeq

\paragraph{Bethe-Vacua $\CS_{\rm BE} (\vec{U},\vec{\nu})$} Bethe-vacua are defined by 
\begin{align}
\begin{split}
\mathcal{S}_{\rm BE}(\vec{U},\vec{\nu}) &:=
\left\{
\vec{z}_0 \;:\; \exp ( \partial_{\vec{Z}} \mathcal{W}_0  (\vec{Z}, \vec{U}, \vec{\nu}) )\big{|}_{\vec{Z}=\log \vec{z}_0} =\vec{1}\,,\,\, {\rm triv.~isotr.}
\right\}
\big/\, \textrm{Weyl}(G)
\\
&= \big{\{}\vec{z}^{(\alpha)} \big{\}}_{\alpha=1}^{|\mathcal{S}_{\rm BE}(\vec{U}, \vec{\nu})|}\;. \label{Bethe vacua}
\end{split}
\end{align}
Here ``triv.\ isotr.'' stands to indicate that Bethe solutions that are invariant under a non-trivial subgroup of the Weyl group should be discarded, since they do not correspond to physical vacua. (See e.g.\ \cite{Closset:2017zgf} for a discussion and references.) We further assume that the background parameters $\vec U$ can be chosen in such a way that all Bethe vacua are massive. Note that the above equations are independent of the 
unphysical shift ambiguities in eq.~\eqref{gtW}. The number of Bethe-vacua at generic $\vec{U}$ is equal to the Witten index \cite{Kim:2010mr,Intriligator2013}
\begin{align}
(\textrm{Witten index})  = |  \mathcal{S}_{\rm BE}|\;.
\end{align}
\paragraph{Perturbative Expansion $\{S_n^{(\alpha,\vec{\ell})}  (\vec{U},\vec{\nu}) \}$}  We consider a formal perturbative expansion of the localization integral
\begin{align}
\begin{split}
&\frac{1}{|\textrm{Weyl}(G)|} \int \frac{    d^{\rG} \delta Z}{(2\pi \hbar)^{\rG/2}}\Upsilon_\hbar\left(\vec U + \left(i \pi +\frac{\hbar}{2}\right) \vec{\nu},\vec Z = \vec{Z}^{(\alpha,\vec{\ell})} +\delta \vec{Z}  \right)  
\\
&\xrightarrow{\quad  \hbar\rightarrow 0 \textrm{\;with fixed }\vec{U}= b \vec{m}  \quad } \mathcal{Z}_{\rm pert}^{(\alpha,\vec\ell)}(\vec{U}, \vec{\nu} ;\hbar)=\exp \left( \sum_{n=0}^\infty \hbar^{n-1} S_n^{(\alpha, \vec{\ell})} (\vec{U},\vec{\nu})\right)
\end{split}
\end{align}
in the limit $\hbar \rightarrow 0$ around a saddle point $\vec{Z} = Z^{(\alpha, \vec{\ell})} (\vec{U},\vec{\nu})$ which satisfies
\begin{align}
\partial_{\vec{Z}} \CW^{\vec{\ell}}_0 (\vec{Z}, \vec{U}, \vec{\nu}) \big{|}_{\vec{Z} = \vec{Z}^{(\alpha,\vec{\ell})}} (\vec{U}, \vec{\nu}) =0\;. \label{saddle point}
\end{align}
For  a given Bethe vacuum $\vec{z}^\alpha \in \CS_{\rm BE}(\vec{U},\vec{\nu})$, there is an  associated logarithmic saddle point $Z^{(\alpha, \vec{\ell})}$ satisfying  $e^{\vec{Z}^{(\alpha,\vec{\ell})}} = \vec{z}^\alpha$ upon a proper choice of shift ambiguities $\vec{\ell}$.  The first two perturbative coefficients $S^{(\alpha, \vec{\ell})}_0$ and $S^{(\alpha, \vec{\ell})}_1$ depend on the choices $\vec{\ell}$ while higher loop coefficients do not. The dependence is of the following form:
\beq
\sum_{n=0}^\infty S^{(\alpha, \vec{\ell})}_n (\vec{U},\vec{\nu}) \hbar^n = \sum_{n=0}^\infty S^{(\alpha, \vec{\ell} = \vec{0})}_n (\vec{U},\vec{\nu}) \hbar^n+2\pi i (U_I\ell^I_u+2\pi i\ell_0)+i\pi\hbar \ell_1\,,\quad \ell^I_u,\,\ell_0,\,\ell_1\in\ZZ\,.\label{gtW-2}
\eeq
The actual localization integral \eqref{Zformula} along the physical cycle $\Gamma_{\mathbb{R}}$ can be asymptotically expanded in terms of the formal perturbative expansion
\begin{align}
\mathcal{Z}_b (\vec{m},\vec{\nu}) \xrightarrow{\qquad b^2 \rightarrow 0 \textrm{ with fixed } \vec{U}=b\vec{m} \qquad} \sum_{\alpha} n_\alpha  \mathcal{Z}_{\rm pert}^{(\alpha,\vec\ell)}(\vec{U}, \vec{\nu} ;\hbar)\big{|}_{\hbar = 2\pi i b^2}\;,
\end{align}
where the integer coefficients $n_\alpha$ and the shift ambiguity $\vec{\ell}$ depends on the direction (phase factor of $b^2$) of the limit. For our purpose (obtaining the $b\rightarrow 1$ expansion), the actual values of them are not relevant and we only need to know the formal perturbative expansion modulo the  ambiguities in eq.~\eqref{gtW-2}.

Now let us give a more explicit formula for the formal  perturbative expansion. For the  perturbative expansions,  we first expand the  integrand around a saddle point~$\vec{Z}^{(\alpha, \vec{\ell})}$,
\begin{align}
\begin{split}
& \exp \bigg{(} \sum_{n=0}^\infty \hbar^{n-1} \mathcal{W}_n(\vec{Z}^{(\alpha, \vec{\ell})} +\hbar^{\frac{1}{2}} \delta \vec{\zeta})  \bigg{)}  
\\
&= \exp \bigg{(}\sum_{n=0}^\infty  \mathcal{W}_n(\vec{Z}^{(\alpha, \vec{\ell})}  )\hbar^{n-1}   - \frac{1}{2}{  \delta \zeta_i     \Pi^{ij} (\vec{Z}^{(\alpha, \vec{\ell})} )  \delta \zeta_j} + \sum_{n=1}^\infty\sum_{\substack{1\le m\le n+2;\\m-n\in 2\ZZ}}\hbar^{n/2} C_{n,m}^{i_1\dots i_m} (\vec{Z}^{(\alpha, \vec{\ell})} )\delta\zeta_{i_1}\dots\delta\zeta_{i_m}\bigg{)}
\\
&= \exp \bigg{(} \sum_{n=0}^\infty  \mathcal{W}_n(\vec{Z}^{(\alpha, \vec{\ell})}  )\hbar^{n-1}  - \frac{1}{2} \delta \zeta_i     \Pi^{ij}  (\vec{Z}^{(\alpha, \vec{\ell})} ) \delta \zeta_j\bigg{)}  
\\
&  \qquad  \times \bigg{(} 1+\sum_{n=1}^\infty    \sum_{\substack{1\le m\le 3n\\ m-n  \in 2\mathbb{Z}}}   \hbar^{\frac{n}2} D_{n,m}^{i_1 \ldots i_m} (\vec{Z}^{(\alpha, \vec{\ell})})  \delta \zeta_{i_1} \ldots \delta \zeta_{i_m}  \bigg{)}\;.\label{expansion of integrand}
\end{split}
\end{align}
In the expansion, the quadratic terms $O(\delta Z^2)$ of the classical part $\mathcal{W}_0$ play the role of the inverse propagator
\begin{align}
\begin{split}
&\Pi^{ij} (\vec{Z}^{(\alpha, \vec{\ell})}) :=- \frac{\partial^2 \mathcal{W}^{\vec{\ell}}_0}{\partial Z_i \partial Z_j}\bigg{|}_{\vec{Z}= \vec{Z}^{(\alpha, \vec{\ell})}} \qquad \textrm{(inverse propagator)} \;,
\end{split}
\end{align}
while other terms play the role of the interaction vertices
\begin{align}
&C_{n,m}^{i_1 \ldots i_m} (\vec{Z}^{(\alpha, \vec{\ell})})   := \frac{1}{m!} \frac{\partial^m  \mathcal{W}^{\vec{\ell}}_{\frac{n-m}2+1} }{\partial Z_{i_1} \ldots \partial Z_{i_m} } \bigg{|}_{\vec{Z}= \vec{Z}^{(\alpha, \vec{\ell})} }\qquad \textrm{(vertices)} \;.
\end{align}
The coefficients $D_{n,m}^{i_1, \ldots, i_n}$ in eq.~\eqref{expansion of integrand} can be written as a finite sum of finite products of the $C$'s. 
Then the  formal expansion $\mathcal{Z}^{(\alpha, \vec{\ell})}_{\rm pert} (\hbar)$ around the saddle point $\vec{Z}^{(\alpha,\vec{\ell}) }$ is given by\footnote{The overall factor $1/\textrm{|\textrm{Weyl}(G)|}$ does not appear here since there are  $\textrm{|\textrm{Weyl}(G)|}$ many saddle points (related by the Weyl group action) of localization integral for each Bethe-vacua, all of which give the same contributions to the asymptotic expansion.}%
\begin{align}\label{Zalpha}
\begin{split}
&\mathcal{Z}^{(\alpha, \vec{\ell})}_{\rm pert}  =\textrm{exp} \bigg{(} \sum_{n=0}^{\infty} \hbar^{n-1} S^{(\alpha, \vec{\ell})}_n \bigg{)}
\\
&=   \exp \bigg{(}\sum_{n=0}^\infty \hbar^{n-1} \mathcal{W}^{\vec{\ell}}_n(\vec{Z}^{(\alpha, \vec{\ell})})\bigg{)}\int \prod_{i=1}^{\rG} \frac{d (\delta \zeta_i)}{\sqrt{2\pi}} \exp \bigg{(} - \frac{1}{2}\delta \zeta_i     \Pi^{ij}  (\vec{Z}^{(\alpha, \vec{\ell})}) \delta \zeta_j \bigg{)}  
\\
&\qquad \qquad  \qquad \qquad \qquad\qquad \qquad   \times \bigg{(} 1+\sum_{n=1}^\infty    \sum_{m =1 }^{3n}   \hbar^{n} D_{2n,2m}^{i_1 \ldots i_{2m}} (\vec{Z}^{(\alpha, \vec{\ell})})  \delta \zeta_{i_1} \ldots \delta \zeta_{i_{2m}}\bigg{)} \;,
\\
& =\exp \bigg{(} - \frac{1}2 \log \det \Pi (\vec{Z}^{(\alpha, \vec{\ell})})  +\sum_{n=0}^\infty \hbar^{n-1} \mathcal{W}^{\vec{\ell}}_n(\vec{Z}^{(\alpha, \vec{\ell})}) \bigg{)} 
\\
& \qquad \qquad \times  \bigg{(}1+  \sum_{n=1}^{\infty} \sum_{m=1}^{3n}\hbar^{n}  D_{2n,2m}^{i_1 \ldots i_{2m}} (\vec{Z}^{(\alpha, \vec{\ell})})  G_{i_1,\ldots, i_{2m}} (\vec{Z}^{(\alpha, \vec{\ell})}) \bigg{)}\;.
\end{split}
\end{align}
To arrive at the last line, we performed the formal Gaussian integrals
\begin{align}
\begin{split}
G_{i_1,\ldots, i_{2m}} &:=\left(\det \Pi\right)^{\frac{1}2}\int \prod_{i=1}^{\rG} \frac{d \delta\zeta_i}{\sqrt{2\pi}}\exp \bigg{(} - \frac{1}{2}\delta\zeta_i     \Pi^{ij}   \delta\zeta_j\bigg{)} \delta\zeta_{i_1}\ldots \delta\zeta_{i_{2m}}
\\
&=\frac{ \partial^{2m} \left( \exp \left( \frac{1}2 \mu^{i} (\Pi^{-1})_{ij} \mu^j \right)  \right)}{\partial \mu^{i_1} \ldots \partial \mu^{i_{2m}} } \bigg{|}_{\mu_i=0}\;.
\\
&
\end{split}
\end{align}
The classical and the one-loop contributions are
\begin{align}
S_0^{(\alpha,\vec\ell)}(\vec U , \vec{\nu}) = \mathcal{W}_0^{\vec\ell} (\vec{Z}^{(\alpha,\vec\ell)})\;, \quad S_1^{(\alpha,\vec\ell)}(\vec U, \vec{\nu}) = \mathcal{W}_1^{\vec\ell} (\vec{Z}^{(\alpha,\vec\ell)}) - \frac{1}2\log\textrm{det}\left(\Pi(\vec{Z}^{(\alpha,\vec\ell)}) \right) \;. \label{S0 and S1 : general}
\end{align}
The terms of higher order in $\hbar$ are defined using $\partial^{\ge 2}\mathcal{W}_0$, $\partial^{\ge 1}\mathcal{W}_1$ and $\partial^{\ge 0}\mathcal{W}_{\ge 2}$ only, and therefore are not affected by the unphysical shifts (\ref{gtW}).

\subsection{\texorpdfstring{Expansion at $b \rightarrow 1$ (Convergent Expansion)}{Expansion at b to1 (Convergent Expansion)}}

Unlike the  $b\rightarrow 0$ limit, we can not use the saddle-point  approximation around $b=1$ since the integrand is smooth around the point. The analytic evaluation of the integral around $b=1$ therefore seems to be intractable as we need to evaluate a generic finite-dimensional integral involving special functions. Most of the interesting physical quantities which can be extracted from the \ptn come from expansion around $b=1$, where the space-time geometry becomes conformally flat. So far, people have heavily relied on numerical approach in  computing the physical quantities.  

However, as already noticed in some literature, the localization integrals are not generic finite-dimensional integral but have several non-trivial hidden structures. One non-trivial property relevant to us is the  factorization property \cite{Pasquetti:2011fj,Beem:2012mb}.  As we will see below,  the property  is so  restrictive that we can determine the  localization integral around $b=1$ from the asymptotic expansion in the $b\rightarrow 0$ limit without  performing actual integration. 

As a main result of this paper, we propose that
\begin{align}
\boxed{
	\begin{array}{lll}
	&\mathcal{Z}_{b} (\vec{m}, \vec{\nu}_0)
	\simeq \sum_{\alpha \in \mathcal{S}_{\rm BE}} \textrm{exp} \bigg{[} \sum_{n=0}^\infty \epsilon^n s_n^{(\a)} (\vec{m}, \vec{\nu}_0) \bigg{]}   \quad (\textrm{for $|\epsilon=1-b|< 1$})\vspace{2mm}
	\\
	&
	\quad\quad\quad\quad:= 
	\sum_{\a \in \mathcal{S}_{\rm BE}} \textrm{exp} \bigg{[} \sum_{n=0}^{\infty} \bigg{(}\hbar_1^{n-1} S^{(\alpha,\vec{\ell})}_n (\vec{U}= b\vec{m},\vec{\nu}_0) + \hbar_2^{n-1} S^{(\alpha,\vec{\ell})}_n (\vec{U}= b^{-1}\vec{m}, \vec{\nu}_0) \bigg{)}\bigg{]} ,  
	\\
	&\textrm{where}
	\\
	&\hbar_1  := 2\pi i (b^2-1) =2\pi i (-2\epsilon+ \epsilon^2)\; , 
	\\
	& \hbar_2  :=2\pi i (b^{-2}-1)  =2\pi i \frac{2\epsilon - \epsilon^2}{(1-\epsilon)^2} =2\pi i (2 \epsilon+3 \epsilon^2 +4\epsilon^3+\ldots ) \;.  
	\end{array}
} \label{main result}
\end{align}
Here $\simeq$ means equality up to an unphysical phase factor
\begin{align}
 \mathcal{Z}_1\simeq \mathcal{Z}_2 \textrm{ if } \mathcal{Z}_1=e^{i \pi \delta}\mathcal{Z}_2 \textrm{ with } \delta \in \mathbb{Q} \;.
\end{align}
The above formula holds only when the R-symmetry mixing parameters $\{\nu_I\}_{I=1}^{r_F}$ in eq.~\eqref{R-symmetry mixing} are cleverly chosen $\vec{\nu} = \vec{\nu}_0$ such that 
\begin{align}
R^{\vec{\nu}_0} + 2 j_3 \in 2 \mathbb{Z}\;, \quad \textrm{for all $1/4$-BPS local operators}  \;. \label{criterion 1}
\end{align}
This guarantees the following condition\footnote{Upon a generic choice of $\vec{\nu}$, the superconformal index contains a term  $q^{\alpha}$ with generic real number $\alpha$. }
\begin{align}
\textrm{only $q^{\rm integer}$ appears in }\CI_q ({\bf m}_I=0,u_I ;\vec{\nu}_0)\; .\label{criterion 2}
\end{align}
Unlike $S_n^{(\alpha, \vec{\ell})}$, the perturbative coefficients $s_n^{(\alpha)}$ are independent on the choices of $\vec{\ell}$ if the  above conditions \eqref{criterion 2} are met. 
Let us give explicit expressions for the first few coefficients $s_n^{(\alpha)}$ in terms of $S_n^{(\alpha,\vec{\ell})}$,
\begin{align}
\begin{split}
s_0^{(\alpha)} (\vec{m},\vec{\nu}_0) &=\frac{i}{2\pi }\left( S_0^{(\alpha, \vec{\ell})}  -  U_I\partial_I S_0^{(\alpha,\vec{\ell})}\right) +2 S_1^{(\alpha,\vec{\ell})} \bigg{|}_{U_I = m_I }\;,\\
 s_1^{(\alpha)} (\vec{m},\vec{\nu}_0)&=0 \;,  \\
s_2^{(\alpha)} (\vec{m},\vec{\nu}_0)& =s_3^{(\alpha)} (\vec{m},\vec{\nu}_0) \\
&= -32 \pi^2 S_3^{(\alpha)} +U_I\partial_I S_1^{(\alpha,\vec{\ell})} + 8\pi i \left(S_2^{(\alpha)}+U_I\partial_I S_2^{(\alpha)}\right)+U_I U_J\partial_{I}\partial_JS_1^{(\alpha,\vec{\ell})} \\
&\qquad -\frac{i}{12\pi}{ U_I U_J U_K} \partial_I\partial_J\partial_KS_0^{(\alpha,\vec{\ell})}  \bigg{|}_{U_I = m_I }\;.\label{s0,s1,s2,s3}
\end{split}
\end{align}
The main result above is valid even for complex $\vec{m}$. This means that if one wants to compute the $\CZ_{b}(\vec{m},\vec{\nu})$ with general choice of R-charge mixing $\vec{\nu}$ which does not satisfy the above condition \eqref{criterion 2}, we can use eq.~\eqref{mixed R-symmetry}
\begin{align}
\CZ_b\left(\vec{m}, \vec{\nu} \right) = \CZ_b \left (\vec{m} +i \pi \left(b+\frac{1}b\right)\left(\vec{\nu} - \vec{\nu}_0\right), \vec{\nu_0} \right) \;, \label{complexified m}
\end{align}
with $\vec{\nu}_0$ satisfying the condition in eq.~\eqref{criterion 2}.

\paragraph{Derivation from Factorization} Our proposal can be derived from the conjectured factorization property of the $S^3_b$ partition function \cite{Pasquetti:2011fj,Beem:2012mb}. This states the following factorization:
\begin{align}
\begin{split}
&\CZ_b (\vec{m},\vec{\nu})  = \sum_{\alpha \in \mathcal{S}_{\rm BE}} B^{(\alpha)} \left(b^2, \vec{u};\vec{\nu} \right)B^{(\alpha)} \left(b^{-2}, \vec{\tilde{u}};\vec{\nu} \right) \textrm{ where}
\\
& u_I:= e^{U_I}:=e^{b m_I }, \;  \tilde{u}_I := e^{\tilde{U}_I}:=e^{b^{-1} m_I}\;.
\end{split}
\label{factorization}
\end{align}
The building block $B^{(\alpha)}$, which is often called the holomorphic block in the literature,
can be considered as the \ptn of the 3d $\mathcal{N}=2$ theory on $\mathbb{R}_{\hbar}^2\times S^1$, with an omega-deformation parameter $\hbar:=2\pi ib^2$ and with an asymptotic boundary condition determined by the Bethe vacuum labelled by $\alpha$:
\begin{align}
\begin{split}
B^{(\alpha)} (b^2, \vec{u};\vec{\nu}) &=  \textrm{Tr}_{\mathcal{H}(\mathbb{R}^2;\alpha)} (-1)^{R^{\vec{\nu}}} q^{\frac{R^{\vec{\nu}}}2 + j_3} \prod_I u_I^{F_I} |_{u_I = e^{bm_I} }\;, \quad q:=e^{2\pi i b^2}\;.
\end{split}
\end{align}
The perturbative expansion of the block $B^{\alpha}(b^2, \vec{u};\vec{\nu})$  in the limit $b^2\rightarrow 0$ with fixed $\vec{u}$  is  expected to be identical to the asymptotic expansion $\mathcal{Z}_{\rm pert}^{(\alpha, \vec{\ell})} (\vec{U},\vec{\nu};\hbar)$ in eq.~\eqref{Zalpha} of the $\mathcal{Z}_b$ in the limit $b^2\rightarrow 0$ with fixed $U_I = b m_I$ around the Bethe vacuum: 
\begin{align}
\begin{split}
&\mathcal{Z}_{\rm pert}^{(\alpha, \vec{\ell})} (\vec{U},\vec{\nu};\hbar) \sim B^{(\alpha)} ( b^2 ,\vec{u}= e^{\vec{U}};\vec{\nu}) 
\\
&\xrightarrow{\quad \;\;    b^2  \rightarrow 0\;\; \textrm{with fixed }U_I = b m_I   \quad  }  \textrm{exp} \bigg{(} \sum_{n=1}^{\infty} \hbar^{n-1} S^{(\alpha,\vec{\ell})}_n (\vec{U}, \vec{\nu})\bigg{)} \bigg{|}_{\hbar = 2\pi i b^2 }\;.
\end{split}
\label{Asymptotic of HB}
\end{align}
Here $\sim$ means equality as an asymptotic expansion in the above limit modulo unphysical factor of the form in eq.~\eqref{gtW-2}.
This is   because of that the anti-holomorphic  part in \eqref{factorization} become trivial  in the asymptotic limit:
\begin{align}
B^{(\alpha)} \left(b^{-2}, \vec{\tilde{u}} = e^{\vec{\tilde{U}}};\vec{\nu} \right) 
\xrightarrow{\quad \;\;   b^2  = -i \delta \rightarrow 0 \;\;  \textrm{with  fixed $U_I= b^2 \tilde{U}_I \in \mathbb{R}$} \quad }  1 + o (e^{-\sharp \frac{2 \pi }\delta })\;.
\end{align}
More precisely, the above triviality  is expected   when  the limit $b^2 \rightarrow 0$ is taken along the direction $b^2 \in - i \mathbb{R}_+$ with  purely imaginary  $\tilde{U}_I$ and the R-charge mixing $\vec{\nu}$ is chosen such that
\begin{align}
B^{(\alpha)} \left(b^{2} = \frac{1}{2\pi i} \log q, \vec{u};\vec{\nu} \right)  = 1+(\ldots) q^{\sharp>0}+(\textrm{higher-order in $q$})\;.
\end{align}
The form of the expansion above is expected when the  R-charge mixing  satisfies  the unitarity constraints from superconformal algebra, such as $R^{\vec{\nu}}(\mathcal{O})\geq \frac{1}2$ for all chiral primary operator $\mathcal{O}$. 
When the condition \eqref{criterion 2} is satisfied, $\vec{\nu} = \vec{\nu}_0$, the block depends  on only $q =e^{2\pi ib^2} $ (instead of $b^2$) and thus
\begin{align}
\begin{split}
\textrm{For $\vec{\nu}_0$ in  eq.~$\eqref{criterion 2}$}\;, \qquad &B^{(\alpha)} (b^2, \vec{u};\vec{\nu}_0)  = B^{(\alpha)} (b^2-1, \vec{u};\vec{\nu}_0) \;,
\\
&B^{(\alpha)} (b^{-2}, \vec{\tilde{u}};\vec{\nu}_0)  = B^{(\alpha)} (b^{-2}-1, \vec{\tilde{u}};\vec{\nu}_0) \;. \label{only q}
\end{split}
\end{align}
Combining eq.~\eqref{factorization}, \eqref{Asymptotic of HB}, \eqref{only q} and  the fact that both of $\hbar_1 = 2\pi i (b^2-1)$ and $\hbar_2 = 2\pi i (b^{-2}-1)$ approach to 0 at  $b\rightarrow 1$, we  derive the expansion in eq.~\eqref{main result}. Since the $b\rightarrow 1$ limit of $\mathcal{Z}_b$ should be smooth,  the final result is valid for any direction in the limit $b\rightarrow 1$.

\paragraph{Comparison with Closset-Kim-Willett \cite{Closset:2017zgf}}  According to \cite{Closset:2017zgf}, the squashed 3-sphere partition function $\mathcal{Z}_b$ at $b=1$ can be written as follows
\begin{align}
\mathcal{Z}_{b=1} (\vec{m},\vec{\nu}) = \mathcal{Z}_{\mathcal{M}_{g=0,p=1}}^{\nu_R =0}  (\vec{m},\vec{\nu}) = \sum_{\alpha \in \mathcal{S}_{\rm BE}} \left(\mathcal{H}^\alpha_{\nu_R = 0} (\vec{m},\vec{\nu})\right)^{-1} \mathcal{F}^{\alpha}_{\nu_R=0}(\vec{m},\vec{\nu})\;.
\end{align}
Here $\mathcal{H}^\alpha_{\nu_R}$ and $\mathcal{F}^\alpha_{\nu_R}$ are so-called `handle-gluing' and `fibering' operators respectively.  These operators serve as basic building blocks of more general supersymmetric partition functions $\mathcal{Z}_{\mathcal{M}_{g,p}}^{\nu_R}$ on  $\mathcal{M}_{g,p}$:
\begin{align}
\begin{split}
&\mathcal{M}_{g,p} := (\textrm{$S^1$-bundle of degree $p$  over a Riemann surface $\Sigma_g$})\;,
\\
&\mathcal{Z}_{\mathcal{M}_{g,p}}^{\nu_R} =  \sum_{\alpha \in \mathcal{S}_{\rm BE}} (\mathcal{H}^\alpha_{\nu_R })^{g-1} (\mathcal{F}^{\alpha}_{\nu_R})^p\;. \label{general Twisted ptns}
\end{split}
\end{align}
For the case when the cycle $[S^1]$ along the fiber is  nontrivial  in $H_1 (\mathcal{M}_{g,p}, \mathbb{Z}_2)$ (i.e.\ $p$ is even), there are  two types of supersymmetric backgrounds depending on the choice of spin-structure along the $S^1$ \cite{Closset:2018ghr}. Depending on the discrete choice, $\nu_R$ can be either $0$ (periodic boundary condition for fermions) or $1/2$ (anti-periodic boundary condition). When $\nu_R=1/2$, we need to turn on the discrete $\mathbb{Z}_2$ holonomy along the $[S^1]$ coupled to a $\U(1)^{\vec{\nu}}_R$ symmetry to preserve some supercharges. 
 When the fiber $[S^1]$ is  trivial in $H_1 (\mathcal{M}_{g,p}, \mathbb{Z}_2)$ (i.e.\ $p$ is odd), only the choice $\nu_R=0$ is allowed. For $\mathcal{M}_{g,p=0} = \Sigma_g \times S^1$ case, for example, the  two different choices correspond to the following  two types of twisted indices
\begin{align}
\begin{split}
&\mathcal{Z}_{\mathcal{M}_{g,p=0}}^{\nu_R = 0} (\vec{m}, \vec{\nu}) = \textrm{Tr}_{\mathcal{H}^{\rm top} (\Sigma_g ;\vec{\nu})}(-1)^{2j_3}  \prod_{I=1}^{r_F} e^{m_I F_I}\;,
\\
&\mathcal{Z}_{\mathcal{M}_{g,p=0}}^{\nu_R=1/2} (\vec{m},\vec{\nu})= \textrm{Tr}_{\mathcal{H}^{\rm top} (\Sigma_g;\vec{\nu})}(-1)^{R^{\vec{\nu}}}  \prod_{I=1}^{r_F} e^{m_I F_I} \;. \label{two types of twisted indices}
\end{split}
\end{align}
Here $\mathcal{H}^{\rm top}(\Sigma_g,\vec{\nu})$ is the Hilbert space of the topologically-twisted 3d $\mathcal{N}=2$ theory on a Riemann surface $\Sigma_g$ of genus $g$. To preserve some supercharges, we turn on the background magnetic flux coupled to the $\U(1)_R^{\vec{\nu}}$ R-symmetry in eq.~\eqref{mixed R-symmetry}.
\begin{align}
\frac{1}{2\pi }\int_{\Sigma_g} dA_{\U(1)_R^{\vec{\nu}}} = (g-1)\;.
\end{align}
Due to the background magnetic flux, the twisted index is well-defined only when the mixing parameters $\vec{\nu}$ are chosen to satisfy following Dirac quantization condition
\begin{align}
R^{\vec{\nu}}  (\mathcal{O}) \times (g-1) \in  \mathbb{Z}\;
\end{align}
for all local operators $\mathcal{O}$ in the 3d theory. 

From the comparison between  eq.~\eqref{S0 and S1 : general} and the explicit expression for $\mathcal{H}^\alpha$  and $\mathcal{F}^\alpha$ given in \cite{Closset:2018ghr}, it is straightforward to check that
\begin{align}
\mathcal{F}^{\alpha}_{\nu_R = 1/2} = \exp \left(i \frac{S^{(\alpha,\vec{\ell})}_0 - U_I \partial_I S^{(\alpha,\vec{\ell})}_0 }{2\pi }\right )\bigg{|}_{U_I = m_I}\;, \quad \mathcal{H}^\alpha_{\nu_R =1/2} = \exp \left(-2S^{(\alpha,\vec{\ell})}_1 \right) \bigg{|}_{U_I = m_I}\;.  \label{Fibering/handle from perturbative invariants}
\end{align}
One simpler way to understand the second equality is using a  factorization of refined twisted index on $S^2$ as explained in \cite{Closset:2018ghr}. The refined index is defined as
\begin{align}
\mathcal{I}^{\rm top}_q (\vec{m},\vec{\nu}) :=\textrm{Tr}_{\mathcal{H}^{\rm top} (\Sigma_g, \vec{\nu})}(-1)^{R^{\vec{\nu}}} q^{\frac{1}2R^{\vec{\nu}}+j_3}  \prod_{I=1}^{r_F} e^{m_I F_I} \;,
\end{align}
and is known to have the following factorization property \cite{Benini:2015noa,Nieri:2015yia}
\begin{align}
\mathcal{I}^{\rm top}_q (\vec{m},\vec{\nu})  = \sum_{\alpha \in \mathcal{S}_{\rm BE}} B^{\alpha} (b^2, \vec{u}, \vec{\nu}) B^{\alpha}(-b^2,\vec{u}, \vec{\nu})\big{|}_{u_I= e^{m_I},\; 2\pi i b^2=\log q }\;.
\end{align}
Taking the $b^2\rightarrow 0$ ($q\rightarrow 1$) limit, we have 
\begin{align} 
\mathcal{I}^{\rm top}_q (\vec{m},\vec{\nu})|_{q=e^{2\pi i b^2}, \, b^2 \rightarrow 0} =   \sum_{\alpha \in \mathcal{S}_{\rm BE}} e^{2S_1^{\alpha} (\vec{m},\vec{\nu})} (1+ o (b^4))\;.
\end{align}
Here we use the asymptotic limit of the holomorphic blocks given in eq.~\eqref{Asymptotic of HB}. In the $q\rightarrow 1$ limit, the refined index becomes $\mathcal{Z}^{\nu_R =\frac{1}2 }_{\mathcal{M}_{p=0,g=0}}$. Comparing the above expression with the general formula  in eq.~\eqref{general Twisted ptns} with $p=g=0$ and $\nu_R =1/2$, we have the second equality in eq.~\eqref{Fibering/handle from perturbative invariants}.  

When the condition in \eqref{criterion 1} is met, there is no difference between two discrete choices, $\nu_R = 0$ or $1/2$, as is obvious in \eqref{two types of twisted indices}. So, in the case, we have
\begin{align}
\begin{split}
\mathcal{Z}_{b=1} &= \sum_{\alpha \in \mathcal{S}_{\rm BE}} (\mathcal{H}^\alpha_{\nu_R = 0})^{-1} \mathcal{F}^{\alpha}_{\nu_R=0}
\\
&= \sum_{\alpha \in \mathcal{S}_{\rm BE}} (\mathcal{H}^\alpha_{\nu_R = 1/2})^{-1} \mathcal{F}^{\alpha}_{\nu_R=1/2} 
\\
&=  \sum_{\alpha \in \mathcal{S}_{\rm BE}}  \exp \left(i \frac{S^{(\alpha,\vec{\ell})}_0 - U_I \partial_I S^{(\alpha,\vec{\ell})}_0 }{2\pi } +2 S_1^{(\alpha, \vec{\ell})}\right ) \;.
\end{split}
\end{align}
It matches the zero-th order approximation, $s_0$ in  eq.~\eqref{s0,s1,s2,s3}, of the general perturbative expansion  \eqref{main result} in $\epsilon = 1-b$.

\subsection{Current and Stress Tensor Correlation Functions} 
We define the free energy and its real part, 
\begin{align}
\begin{split}
\mathcal{F}_b (\vec{m}, \vec{\nu}) &= -\log \mathcal{Z}_b (\vec{m},\vec{\nu}) = - \log \mathcal{Z}_b \bigg{(} \vec{m}+ i \pi \left(b+ \frac{1}{b} \right) (\vec{\nu}- \vec{\nu}_0), \vec{\nu}_0 \bigg{)}
\\
&= -\log \sum_{\alpha} \exp \bigg{(}\sum_{n=0}^\infty s_n \bigg{(} \vec{m}+ i \pi \left(b+ \frac{1}{b} \right) (\vec{\nu}- \vec{\nu}_0), \vec{\nu}_0 \bigg{)} (1-b)^n \bigg{)}\;,
\\
F_b (\vec{m},\vec{\nu})&:= \Re[\mathcal{F}_b(\vec{m},\vec{\nu})]\;.
\end{split}
\end{align}
Here $\vec{\nu}_0$ is a choice of R-charge mixing satisfying \eqref{criterion 1}. 
The infra-red (IR) superconformal $R$-charge  $R^{\vec{\nu}_{\rm IR}}$ can be  determined by the F-maximization principle \cite{Jafferis:2010un} which states that\footnote{This only works when all the Cartan subalgebra of the IR flavor symmetry comes from the UV flavor symmetry $G_F$.  }
\begin{align}
	\textrm{$F_{b=1}(\vec{m}=\vec{0},\vec{\nu} )$ is maximized at $\vec{\nu}=\vec{\nu}_{\rm IR}$}\;. \label{F-maximization}
\end{align}
Then, basic CFT data at the IR fixed point can be obtained as follows \cite{Closset:2012vg,Closset:2012ru}
\begin{align}
\begin{split}
F &= F_{b=1} \big{(}\vec{m}=\vec{0}, \vec{\nu}_{\rm IR}\big{)} \;,
\\
&= - \Re \bigg{[}\log \sum_{\alpha} \exp(s_0^{(\alpha)} \big{(}\vec{m},\vec{\nu}_0 \big{)} )\bigg{]} \bigg{|}_{\vec{m} = 2\pi i (\vec{\nu}_{\rm IR} - \vec{\nu}_0)}\;,
\\
C_{IJ} &= 8\, \partial_{m_I} \partial_{m_J} F_{b=1}\bigg{|}_{\vec{m}=\vec{0}, \vec{\nu}= \vec{\nu}_{\rm IR}}
\\
=&-8 \Re\bigg{[}    \frac{\sum_{\alpha \in \mathcal{S}_{\rm BE}} \textrm{exp}\left( s_0^{(\alpha)}(\vec{m}, \vec{\nu}_0 )  \right) \left(  \partial_{m_I} \partial_{m_J}s_0^{(\alpha)}(\vec{m}, \vec{\nu}_0 )    \right)}{\sum_{\alpha \in \mathcal{S}_{\rm BE}} \textrm{exp}\left( s_0^{(\alpha)}(\vec{m}, \vec{\nu}_0 )  \right)} \bigg{]}_{\vec{m}= 2\pi  i (\vec{\nu}_{\rm IR} - \vec{\nu}_0)}\;,
\\
C_T&=\frac{8}{\pi^2} \partial_{b} \partial_{b} F_b\big{|}_{\vec{m}= \vec{0}, \vec{\nu} = \vec{\nu}_{\rm IR},b=1} \\
&= -\frac{8}{\pi^2  } \Re\bigg{[} \frac{\sum_{\alpha \in \mathcal{S}_{\rm BE}} \textrm{exp}\left( s_0^{(\alpha)}(\vec{m}, \vec{\nu}_0 )  \right) \left( 2s_2^{(\alpha)}(\vec{m}, \vec{\nu}_0 ) + m_I \partial_{m_I}s_0^{(\alpha)}(\vec{m}, \vec{\nu}_0 )    \right)}{\sum_{\alpha \in \mathcal{S}_{\rm BE}} \textrm{exp}\left( s_0^{(\alpha)}(\vec{m}, \vec{\nu}_0 )  \right) } \bigg{]}_{\vec{m}= 2\pi  i (\vec{\nu}_{\rm IR} - \vec{\nu}_0)}  \;. \label{Z, k and CT}
\end{split} 
\end{align}
The central charges $C_{IJ}$  and  $C_T$ are defined in eq.~\eqref{CT_def} and \eqref{CJJ_def}.
We fix the normalization of $T_{\mu\nu}$ and $J_\mu$ such that
\begin{align}
\begin{split}
&\left( C_T \textrm{ of a free theory of single chiral $\Phi$}  \right)=1\;, 
\\
&\left(C_{JJ} \textrm{ for the $\U(1)$ symmetry of  a free chiral theory under which $\Phi$ has charge $+1$} \right) =1\;. \label{Normalization of CT}
\end{split}
\end{align}

\section{Examples} \label{sec:examples}

In this section we present some concrete computations of the $b\rightarrow 1$ expansion \eqref{main result}. 
In addition to analytical computation as presented above,
we independently confirm  the computations by  numerically evaluating the coefficients directly from the localization integral. Using the expansion, we present analytic expression of the stress-energy tensor central charge, $C_T$, for candidates for minimal (with lowest $C_T$)  3d $\mathcal{N}=2$ SCFTs. It include the  candidates for minimal 3d $\mathcal{N}=2$ theory \cite{Bobev:2015vsa}, minimal 3d $\mathcal{N}=4$ theory \cite{Gang:2018huc} and minimal 3d $\mathcal{N}=2$ theory \cite{Gang:2017lsr,Gang:2018wek,Gaiotto:2018yjh,Benini:2018bhk}  with $\SU(3)$ flavor symmetry. 
  
\subsection{Free Chiral Multiplet and Critical Wess-Zumino model} 
The $S^3_b$-\ptn and the superconformal index for the theory of an $\mathcal{N}=2$ free chiral multiplet are 
\begin{align}
\begin{split}
&\mathcal{Z}_b (m,\nu):=\psi_\hbar \left(U+ \left(i \pi+ \frac{\hbar}2 \right) \nu \right)\bigg{|}_{U= b m}\;,
\\
&\mathcal{I}_q ({\bf m},u ;q) = \mathcal{I}_{\Delta}\big{(}{\bf m},u(-q^{1/2})^\nu;q \big{)}\;.
\end{split}
\end{align}
The theory has a $\U(1)_\Phi$ flavor symmetry under which $\Phi$ has charge $+1$. The parameters $m$ and $\nu$ are the real mass parameter for the flavor symmetry  and the R-symmetry mixing parameter $R^{\nu}= R^{\nu=0} + \nu \U(1)_\Phi$, respectively, and $({\bf m},u)$ is a pair (background magnetic flux, fugacity) for the  flavor symmetry. 
\begin{table}[htbp]
	\centering
	\begin{tabular}{|l|l|l|}
		\hline
		& $\phi$ & $\psi$    \\ \hline
	$\U(1)_R^{\nu=0}$	& $0$ & $-1$     \\ \hline
	$\U(1)_{\Phi}$	&  $1$ &  $1$     \\ \hline
	\end{tabular}
    \caption{$\U(1)_R \times \U(1)_\Phi$ symmetry of the free chiral theory. $\phi$ and $\psi$ are the complex scalar and the fermion field inside the $\mathcal{N}=2$ chiral multiplet respectively. }
\end{table}
The condition \eqref{criterion 2} is satisfied when 
\begin{align}
\nu_0  \in 2\mathbb{Z}\;. \label{conditions on nu}
\end{align}
In this example, we do not need to solve the saddle point equations to obtain an asymptotic expansion in $b\rightarrow 0$ limit since there is no integration.\footnote{According to a 3d $\mathcal{N}=2$ mirror duality \cite{Aharony:1997bx}, as we will see below, the theory is dual to a  theory  with a $\U(1)$ vector multiplet coupled to a single chiral multiplet. In the dual description, there is a single Bethe vacuum and the asymptotic expansion here can be thought as the expansion around the unique vacuum in the dual description. } The asymptotic expansion is given in  \eqref{asymptotic of QDL}, 
\begin{align}
\begin{split}
&\sum_{n=0}^\infty \hbar^{n-1}  S^{\vec{\ell}}_n (U, \nu) = \sum_{n=0}^\infty \frac{B_n}{n!} \hbar^{n-1} \textrm{Li}_{2-n}(e^{-U +(i \pi + \frac{\hbar}2) \nu})+ \frac{1}\hbar ( 4\pi^2 \ell_0 + 2\pi i \ell_u U+ i \pi \hbar \ell_1) \;,
\\
& \quad \Rightarrow  S^{\vec{\ell}}_0 (U,\nu) = \textrm{Li}_2(e^{-U -i \pi \nu}) + 4\pi^2 \ell_0 + 2\pi i \ell_u U  \;,
\\
& \quad\quad  \; \; S^{\vec{\ell}}_1 (U,\nu) = -\frac{1}2 \log (1-e^{-U-i\pi \nu}) + \frac{1}2 \nu \log (1-e^{-U-i \pi \nu})+ \pi i \ell_1 \;,
\\
&\quad \quad\;\; S_2 (U,\nu)= \frac{2-6 \nu +3 \nu^2}{24 (e^{U+i \pi \nu}-1)}\;, \quad S_3 (U,\nu) =\frac{e^{U+i \pi \nu} \nu (2-3\nu+\nu^2)}{48 (1-e^{U+i \pi \nu})^2}\;.
\end{split}
\end{align}
Then using eq.~\eqref{s0,s1,s2,s3}, we  have 
\begin{align}
\begin{split}
&\exp \big{(}s_0 (m,\nu) \big{)}= \exp \left[ \frac{i}{2\pi} \left( (2\pi i(1-\nu) -  m) \log (1-e^{-m-i \pi \nu}) + \textrm{Li}_2 (e^{-m- i \pi \nu}) \right) \right]\;,
\\
&s_2 (m,\nu ) =s_3 (m,\nu )
\\
&= \frac{e^{m+i \pi  \nu } \left[i m^3-6 \pi  (\nu -1) m^2+4 \pi ^2 (2 \pi  \nu ^3+3 i \nu ^2-6 \pi  \nu ^2-6 i \nu +4 \pi  \nu +2 i)\right]}{12 \pi  \left(-1+e^{m+i \pi  \nu }\right)^2}
\\
&\quad +\frac{ m e^{m+i \pi  \nu }  (-6 i \pi  \nu ^2+12 i \pi  \nu +3 \nu -4 i \pi -3) +  (-6 i \pi  \nu ^2-3 m \nu +12 i \pi  \nu +3 m-4 i \pi )}{6 \left(-1+e^{m+i \pi  \nu }\right)^2}\;.
 \label{s0 and s2 for free chiral}
\end{split}
\end{align}
Note that the $\exp (s_0)$ is independent on the choice of $\ell_u \in \mathbb{Z}$.
We finally have $(\nu_0 \in 2\mathbb{Z})$
\begin{align}
\begin{split}
&\log \mathcal{Z}_b (m,\nu ) = \log \mathcal{Z}_b \left(m + i \pi (b+b^{-1}) (\nu- \nu_0), \nu_0 \right)
\\
&=s_0 \big{(}m+2\pi i (\nu- \nu_0),\nu_0 \big{)}+ \big{[}s_2 \big{(}m+2\pi i (\nu- \nu_0),\nu_0)\big{)}
\\
& \qquad +i \pi (\nu-\nu_0)s^{(1,0)}_0 \big{(}m+2\pi i (\nu- \nu_0),\nu_0\big{)} \big{]}\big{(}(1-b)^2+(1-b)^3 \big{)} +o\big{(}(b-1)^4\big{)}\;.
\end{split}
\end{align}
Here
\begin{align}
s^{(1,0)}_0 (m, \nu) := \partial_m s_0 (m,\nu) = \frac{i m-2 \pi  \nu +2 \pi }{2 \pi -2 \pi  e^{m+i \pi  \nu }} \;.
\end{align}
As a consistency check, the expression is actually independent on the choice of $\nu_0 \in 2\mathbb{Z}$.
For a free chiral theory, the superconformal IR R-symmetry corresponds to $\U(1)^{\nu}_R$  with $\nu= \frac{1}2$ and the stress-energy tensor central charge  is 
\begin{align}
\begin{split}
(\textrm{$C_T$ of a free $\Phi$}) &= - \frac{8}{\pi^2} \Re \partial_b^2\left[ \log \mathcal{Z}_b \left(m=0,\nu = \frac{1}2 \right)\right]_{b=1} 
\\
& = - \frac{8}{\pi^2}\Re  \partial_b^2\left[\log  \mathcal{Z}_b \left(m=\frac{i \pi }2 (b+b^{-1}), \nu_0=0 \right)\right]_{b=1}
\\
&= - \frac{16}{\pi^2}\Re\left[ s_2(i \pi, \nu_0=0)+ \frac{i\pi}2 s_0^{(1,0)} (i \pi, \nu_0=0) \right]=1\;.
\end{split}
\end{align}
In the second line, we use eq.~\eqref{complexified m}. The $C_{JJ}$ for the flavor $\U(1)_\Phi$ symmetry is
\begin{align}
\begin{split}
C_{JJ} &= - 8 \Re \partial_m^2\left[ \log \mathcal{Z}_b \left(m,\nu = \frac{1}2 \right)\right]_{m=0,b=1}
\\
&= - 8 \Re \partial_m^2\left[ \log \mathcal{Z}_b \left(m+\frac{i \pi}2 (b+b^{-1}),\nu_0 = 0  \right)\right]_{m=0,b=1}
\\ 
&= -8 \Re \partial_m^2 s_0(m+\pi i ,\nu_0 = 0)|_{m=0} = - 8\Re   \frac{e^m (i m-i+2 \pi )+i}{2 \pi  \left(e^m-1\right)^2}\big{|}_{m=i \pi}
\\
&= 1 \;.
\end{split}
\end{align}
From the above computations for the free chiral theory, we confirmed the normalization in eq.~\eqref{Normalization of CT}. 

On the other hand, the IR R-symmetry of the critical Wess-Zumino model (cWZ, a chiral multiplet $\Phi$ with superpotential $W_{\rm sup} = \Phi^3$) corresponds to $\U(1)^{\nu}_R$ with $\nu=2/3$ and 
\begin{align}
\begin{split}
(\textrm{$C_T$ of cWZ})   &= - \frac{8}{\pi^2} \Re \partial_b^2\left[ \log \mathcal{Z}_b \left(m=0, \nu=\frac{2}3\right)\right]
\\
&= - \frac{8}{\pi^2} \Re \partial_b^2\left[ \log \mathcal{Z}_b \left(m=\frac{ 2\pi i }3 (b+b^{-1}), \nu_0=0\right)\right] 
\\
&= - \frac{16}{\pi^2} \Re \left[ s_2 \left(\frac{4 i \pi}3 , \nu_0=0 \right)+ \frac{2i\pi}3 s_0^{(1,0)} \left(\frac{4 i \pi}3 , \nu_0=0\right) \right]
\\
&= \frac{16}{243} \left(16-\frac{9 \sqrt{3}}{\pi }\right)\simeq  0.726785\;. 
\end{split}
\end{align}
The  result nicely matches  the analytic result in \cite{Witczak-Krempa:2015jca} obtained from direct integration using eq.~\eqref{QDL : integral}. We reproduce the result from a drastically simpler computation.  

\subsection{\texorpdfstring{$\U(1)_k $ Coupled to a Chiral Multiplet of Charge $+1$}{U(1)k Coupled to a Chiral Multiplet of Charge +1}} The squashed 3-sphere partition function and the superconformal index of theory are
\begin{align}
\begin{split}
&\mathcal{Z}_{b} \big{(}m,\nu \big{)} = \int \! \frac{dZ}{\sqrt{2\pi \hbar}}\, e^{\frac{(k+\frac{1}2) Z^2- 2 Z(U+ (i \pi  + \frac{\hbar}2) \nu )   }{2\hbar}} \psi_{\hbar} (Z) \bigg{|}_{U= bm }\; ,
\\
&\CI_q ({\bf m},u;\nu) = \sum_{{\bf n} \in \mathbb{Z}} \oint_{|v|=1} \frac{dv}{2\pi i v} \,v^{-\bf m} ((-q^{\frac{1}{2}})^{\nu}u )^{-\bf n} v^{(k+\frac{1}{2}){\bf n}} \CI_{\Delta}({\bf n},v;q)\;.
\end{split}
\end{align}
The theory has a topological $\U(1)_J$ flavor symmetry associated to the Abelian gauge symmetry. As in the previous example, $m$ and $\nu$ are the real mass parameter and the R-symmetry mixing parameter for the flavor symmetry respectively, and $({\bf m},u)$ are (background magnetic flux, fugacity) for the flavor symmetry. The proper Chern-Simons level quantization of the theory is \cite{Redlich:1983dv,Niemi:1983rq}
\begin{align}
k \in \mathbb{Z}+\frac{1}{2} \;.
\end{align}
The condition \eqref{criterion 2} is satisfied when 
\begin{align}
\nu_0 + \left(k+\frac{1}{2}\right) \in 2\mathbb{Z}\;. \label{conditions on nu and k}
\end{align}
The Bethe-vacua of the theory are determined by the following algebraic equations
\begin{align}
\CS_{\rm BE} (U,\nu) = \left\{z\;: \;z^{k+\frac{1}{2}}\left(1-\frac{1}{z}\right) =e^U (-1)^\nu \right\}\;. \label{Bethe-vacua for u(1)+Phi}
\end{align}
Note that, for generic $U$
\begin{align}
|\CS_{\rm BE}| = |k|+\frac{1}2\;,
\end{align}
which  matches the Witten index computation in \cite{Intriligator:2013lca}. The $b\rightarrow 0$ perturbative expansion coefficients $\{S_n^{(\alpha, \vec{\ell})} (U,\nu)\}_{n=0}^\infty$ can be computed using the method summarized in the section \ref{sec: b goes to 0}.  Up to 3-loop, the perturbative invariants are as follows
\begin{align}
\begin{split}
&S^{(\alpha, \vec{\ell})}_0 (U, \nu )=  \textrm{Li}_2 (e^{-Z} )+4 \pi^2 \ell_0+ 2\pi i\ell_z Z  - (U+i \pi \nu)Z + \frac{k+1/2}2 Z^2 \bigg{|}_{Z=Z^{(\alpha, \vec{\ell})}(U,\nu)}\;,
\\
&e^{2 S^{(\alpha, \vec{\ell})}_1 (U,\nu)} =  \frac{2z^{1-\nu}}{A}  \bigg{|}_{z=\exp (Z^{(\alpha, \vec{\ell})}(U,\nu))}\;,
\\
&S_2^{\alpha} (U,\nu=0)=\frac{ A^2 (2 k-5)+A (-14 k+6 z+17)+10 (2 k-2 z-1)}{12 A^3}  \bigg{|}_{z=\exp \left(Z^{(\alpha, \vec{\ell})}(U,\nu)\right)}\;,
\\
&S_2^{\alpha} (U,\nu=-1)=\frac{A^2 (2 k-3 z+4)+A (-14 k-6 z+17)+10 (2 k-2 z-1)}{12 A^3}\bigg{|}_{z=\exp \left(Z^{(\alpha, \vec{\ell})}(U,\nu)\right)} \;,
\\
&S_3^\alpha(U,\nu=0) 
\\
&=\frac{(4k^2-1)(1-z)z\big{(}A^3+A^2 (6 k-z-15)+10A (2 z-4 k+5)+30 (2 k-2 z-1)\big{)} }{12 A^6}\bigg{|}_{z=\exp \left(Z^{(\alpha, \vec{\ell})}(U,\nu)\right)}\;,
\\
&S_3^\alpha(U,\nu=-1) =\frac{(2k-1)(1-z) z  \big{(}A^2 k (6 k+1)+A k (20-40 k)+15 (1-2 k)^2 \big{)}}{6 A^6}\bigg{|}_{z=\exp \left(Z^{(\alpha, \vec{\ell})}(U,\nu)\right)}\;.
\end{split}
\end{align}
Here we define
\begin{align}
A:=2k(z-1)+z+1\;.\label{def : A}
\end{align}
The $Z^{(\alpha, \vec{\ell})}$ and $\ell_z^\alpha \in \mathbb{Z}$ is chosen such that
\begin{align}
\partial_Z \CW^{\vec{\ell}}_0 \big{|}_{Z = Z^{(\alpha, \vec{\ell})}} = \left(k+\frac{1}2\right) Z  +\log (1-e^{-Z}) -  U - i \pi \nu + 2\pi i \ell_z^\alpha  \big{|}_{Z = Z^{(\alpha, \vec{\ell})}}  =0\;.
\end{align}
For given $z^\alpha \in \CS_{\rm BE} (U,\nu)$ in eq.~\eqref{Bethe-vacua for u(1)+Phi}, the choice of $(Z^\alpha = \log z^\alpha , \ell_z^\alpha)$ is not unique but has following shift ambiguity
\begin{align}
Z^\alpha \rightarrow Z^\alpha + 2\pi i t \;, \quad \ell_z^\alpha \rightarrow \ell_z^\alpha - \left(k+\frac{1}2\right) t\;, \quad t\in \mathbb{Z}\;. \label{shift-ambiguity for u(1)+Phi}
\end{align}
Then, the $b\rightarrow 1$  expansion coefficients $s_n^{\alpha}(m,\nu)$ in eq.~\eqref{s0,s1,s2,s3} up to $n=3$ are 
\begin{align}
\begin{split}
 &\exp \left(s^{\alpha}_0 (m, \nu) \right)= \frac{2 \exp \left(\frac{i \text{Li}_2\left(e^{-Z}\right)}{2 \pi } + \frac{i ( k+\frac{1}2) Z^2}{4 \pi }-\frac{\nu }{2} Z +(1-\ell_z^\alpha) Z  \right) }{A}\bigg{|}_{ Z = Z^{(\alpha, \vec{\ell})}(U=m, \nu)}\;,
\\
&s^\alpha_2 (m,\nu=0) = \frac{1}{3\pi A^6} \bigg{(}A^5 \left(4 i \pi ^2 k-3 \pi  m-10 i \pi ^2\right)
\\
&+A^4 \left(-28 i \pi ^2 k+6 \pi  m^2 z+6 \pi  m z+12 i \pi ^2 z+34 i \pi ^2\right)
\\
&+A^3 \big{(}-16 i \pi ^2 k^2 m z^2+16 i \pi ^2 k^2 m z+32 \pi ^3 k^2 z^2-32 \pi ^3 k^2 z+32 i \pi ^2 k m z^2-32 i \pi ^2 k m z
\\
&\qquad \quad  -16 \pi ^3 k z^2+16 \pi ^3 k z+40 i \pi ^2 k+2 i m^3 z^2-2 i m^3 z-12 \pi  m^2 z^2-24 \pi  m^2 z
\\
&\qquad \quad  -4 i \pi ^2 m z^2+4 i \pi ^2 m z-40 i \pi ^2 z-20 i \pi ^2\big{)}
\\
&+A^2 \big{(}192 \pi ^3 k^3 z^2-192 \pi ^3 k^3 z+224 i \pi ^2 k^2 m z^2-224 i \pi ^2 k^2 m z-512 \pi ^3 k^2 z^2+512 \pi ^3 k^2 z
\\
&\qquad \quad -256 i \pi ^2 k m z^2+256 i \pi ^2 k m z+304 \pi ^3 k z^2-304 \pi ^3 k z+48 \pi  m^2 z^2
\\
&\qquad \quad +72 i \pi ^2 m z^2-72 i \pi ^2 m z-48 \pi ^3 z^2+48 \pi ^3 z \big{)}
\\
& +A \big{(}-1280 \pi ^3 k^3 z^2+1280 \pi ^3 k^3 z-480 i \pi ^2 k^2 m z^2+480 i \pi ^2 k^2 m z +2240 \pi ^3 k^2 z^2
\\
&\qquad \quad-2240 \pi ^3 k^2 z+480 i \pi ^2 k m z^2-480 i \pi ^2 k m z-1280 \pi ^3 k z^2
\\
&\qquad \quad +1280 \pi ^3 k z-120 i \pi ^2 m z^2+120 i \pi ^2 m z+240 \pi ^3 z^2-240 \pi ^3 z \big{)}
\\
&+1920 \pi ^3 k^3 z^2-1920 \pi ^3 k^3 z-2880 \pi ^3 k^2 z^2+2880 \pi ^3 k^2 z+1440 \pi ^3 k z^2
\\
&-1440 \pi ^3 k z-240 \pi ^3 z^2+240 \pi ^3 z\bigg{)}\bigg{|}_{ z = \exp\left(Z^{(\alpha, \vec{\ell})}(U=m, \nu)\right)}\;,
\\
&s^\alpha_2 (m,\nu=-1) =\frac{1}{3\pi A^6} \bigg{(}A^5 \left(4 i \pi ^2 k+3 \pi  m z-6 \pi  m-6 i \pi ^2 z+8 i \pi ^2\right)
\\
&+A^4 \left(-28 i \pi ^2 k+6 \pi  m^2 z+6 \pi  m z-12 i \pi ^2 z+34 i \pi ^2\right)
\\
&+A^3 \big{(}-16 i \pi ^2 k^2 m z^2+16 i \pi ^2 k^2 m z-16 i \pi ^2 k m z^2+16 i \pi ^2 k m z +40 i \pi ^2 k+2 i m^3 z^2-2 i m^3 z
\\
&\qquad \quad -36 \pi  m^2 z-4 i \pi ^2 m z^2+4 i \pi ^2 m z-40 i \pi ^2 z-20 i \pi ^2\big{)}
\\
&+A^2 \big{(}192 \pi ^3 k^3 z^2-192 \pi ^3 k^3 z+224 i \pi ^2 k^2 m z^2-224 i \pi ^2 k^2 m z-64 \pi ^3 k^2 z^2+64 \pi ^3 k^2 z
\\
&\qquad \quad -64 i \pi ^2 k m z^2+64 i \pi ^2 k m z-16 \pi ^3 k z^2+16 \pi ^3 k z+48 \pi  m^2 z^2-24 i \pi ^2 m z^2+24 i \pi ^2 m z \big{)}
\\
&+A \big{(}-1280 \pi ^3 k^3 z^2+1280 \pi ^3 k^3 z-480 i \pi ^2 k^2 m z^2+480 i \pi ^2 k^2 m z+1280 \pi ^3 k^2 z^2-1280 \pi ^3 k^2 z
\\
&\qquad \quad +480 i \pi ^2 k m z^2-480 i \pi ^2 k m z-320 \pi ^3 k z^2+320 \pi ^3 k z-120 i \pi ^2 m z^2+120 i \pi ^2 m z \big{)}
\\
& +1920 \pi ^3 k^3 z^2-1920 \pi ^3 k^3 z-2880 \pi ^3 k^2 z^2+2880 \pi ^3 k^2 z+1440 \pi ^3 k z^2
\\
&-1440 \pi ^3 k z-240 \pi ^3 z^2+240 \pi ^3 z  \bigg{)}\bigg{|}_{ z = \exp\left(Z^{(\alpha, \vec{\ell})}(U=m, \nu)\right)}  \;. \label{s0,s2 for u(1) +Phi}
\end{split}
\end{align}
Here $A$  is defined in eq.~\eqref{def : A} and we use the followings
\begin{align}
\begin{split}
&\partial_U Z^{(\alpha , \vec{\ell})}(U,\nu) = \frac{2 (z-1)}{A} \bigg{|}_{ z = \exp\left(Z^{(\alpha, \vec{\ell})}\right)}\;,
\\
&\partial_U  \partial_U  Z^{(\alpha , \vec{\ell})}(U,\nu) =\frac{8 (z-1) z}{A^3}\bigg{|}_{ z = \exp\left(Z^{(\alpha, \vec{\ell})}\right)}\;,
\\
&\partial_U  \partial_U \partial_U Z^{(\alpha , \vec{\ell})}(U,\nu) =-\frac{16 (z-1) z (A z+A-6 z)}{A^5}\bigg{|}_{ z = \exp\left(Z^{(\alpha, \vec{\ell})}\right)}\;.
\end{split}
\end{align}
Under the shift \eqref{shift-ambiguity for u(1)+Phi}, the expression $\exp \left( s^\alpha_0 (m, \nu) \right)$ transforms as
\begin{align}
\exp \left(s^\alpha_0 (m,\nu) \right) \rightarrow e^{i \pi t (\nu - (k+\frac{1}2)t )} \exp \left(s^\alpha_0 (m,\nu) \right)\;.
\end{align}
Thanks to the condition in eq.~\eqref{conditions on nu and k}, the phase factor is just $1$ and the $\exp \big{(}s^\alpha_0(m,\nu) \big{)}$ is invariant under the shift.  So, we finally have
\begin{align}
\begin{split}
&\log \mathcal{Z}_b \left(m,\nu \right)
\\
&=\log \mathcal{Z}_b\left(m+i \pi (b+b^{-1})(\nu -\nu_0), \nu_0 \right)
\\
&=\log \sum_\alpha  \exp \bigg{[}s^\alpha_0 \big{(}m+2\pi i (\nu-\nu_0),\nu_0\big{)} +\left( (1-b)^2+(1-b)^3 \right) \times \bigg{(} s^{\alpha}_2\big{(}m+2\pi i (\nu-\nu_0), \nu_0 \big{)}
\\
&\qquad \qquad \qquad \;\; +\pi i (\nu-\nu_0) s_0^{\alpha,(1,0)}\big{(}m+2\pi i (\nu-\nu_0), \nu_0 \big{)} \bigg{)} +o \left((1-b)^4\right) \bigg{]}\;,
\end{split}
\end{align}
where $\nu_0$ is chosen as in eq.~\eqref{Bethe-vacua for u(1)+Phi}.
Here we define
\begin{align}
\begin{split}
&s_0^{\alpha,(1,0)} (m,\nu):= \partial_m s^\alpha (m,\nu) 
\\
&=\frac{i A m (z-1)-2 \pi  A (\nu  (z-1)+1)+4 \pi  z}{\pi  A^2}\bigg{|}_{ z = \exp\left(Z^{(\alpha, \vec{\ell})}\right)}   \;,
\\
&s_0^{\alpha,(2,0)} (m,\nu):= \partial_m \partial_m s^\alpha (m,\nu) 
\\
&=\frac{i A^3 (z-1)+4 \pi  A^2 z+4 i A m (z-1) z-8 \pi  A z (-\nu +\nu  z+z+2)+32 \pi  z^2}{\pi  A^4}\bigg{|}_{ z = \exp\left(Z^{(\alpha, \vec{\ell})}\right)} \;.
\label{partial s0 for u(1) +Phi}
\end{split}
\end{align}

\paragraph{$k=-1/2$ Case : Dual to Free Chiral Multiplet} 
In the case, the squashed three-sphere partition function is  
\begin{align}
&\mathcal{Z}_{b} \big{(}m,\nu=0 \big{)} = \int \! \frac{dZ}{\sqrt{2\pi \hbar}} \, e^{-\frac{ ZU   }{\hbar}} \psi_{\hbar} (Z)\bigg{|}_{U=bm}\;.
\end{align}
Recall that $\hbar:=2\pi ib^2$. The logarithmic Bethe-vacua equation is 
\begin{align}
\begin{split}
&\log (1-e^{-Z}) -m + 2\pi i \ell_z =0\;,
\\
&\quad\Rightarrow Z =-\log  (1-e^m )\;, \quad \ell_z =0\;. 
\end{split}
\end{align}
From the  computations in eq.~\eqref{s0,s2 for u(1) +Phi} (we choose $\nu_0=0$),
\begin{align}
\begin{split}
&\exp \left(s_0(m,\nu_0=0) \right) = - \exp \left[\frac{i}{2\pi}\textrm{Li}_2(e^{-Z}) + Z \right]= - \exp \left[\frac{i}{2\pi}\textrm{Li}_2(1-e^m) -\log (1-e^m) \right]
\\
&= \exp \left[\frac{i}{2\pi } \left(\textrm{Li}_2 (e^{-m})  +(2\pi i - m) \log (1-e^{-m}) - \frac{1}2m^2 +2\pi i m -\frac{ \pi^2 }6\right)\right]\;,
\\
&s_2 (m,\nu_0 =0) =\frac{i \left(m^3 (z-1) z-6 i \pi  m^2 (z-1) z-2 \pi  m (z-1) (4 \pi  z+3 i)+4 \pi ^2 (1-2 z)\right)}{12 \pi }
\\
&= \frac{e^m m \left(i m^2+6 \pi  m+2 \pi  (3-4 i \pi )\right)+2 i \pi  e^{2 m} (2 \pi +3 i m)-4 i \pi ^2}{12 \pi  \left(e^m-1\right)^2}\;.
\end{split}
\end{align}
Here we use following identity
\begin{align}
\textrm{Li}_2 (1-u)= \textrm{Li}_2 \left(\frac{1}{u}\right) + \frac{\pi^2}3 +\frac{1}2 \log^2 (-u)  - \log u \log (1-u) \;.
\end{align}
Comparing with $s_0$ and $s_1$ for a free chiral theory in eq.~\eqref{s0 and s2 for free chiral},
\begin{align}
\begin{split}
&s^{\U(1)_{-\frac{1}{2}}+\Phi}_0(m, \nu_0=0) = s^{ \Phi}_0(m, \nu_0=0) + \frac{m^2}{4\pi i} -m - \frac{i \pi}{12}\;,
\\
&s^{\U(1)_{-\frac{1}{2}}+\Phi}_2(m, \nu_0=0) = s^{ \Phi}_2(m, \nu_0=0) -\frac{m}2 + \frac{i \pi}3\;.
\end{split}
\end{align}
This is compatible with following identity ($ \mathcal{Z}_1\simeq \mathcal{Z}_2$ means $\mathcal{Z}_1=e^{i \pi \delta}\mathcal{Z}_2$ with $\delta \in \mathbb{Q}$)
\begin{align}
\begin{split}
 &\int \! \frac{dZ}{\sqrt{2\pi \hbar}} \, e^{-\frac{ ZU   }{\hbar}} \psi_{\hbar} (Z) \simeq e^{\frac{U^2- (2\pi i +\hbar )U}{2\hbar} + \frac{i \pi (b^2+b^{-2})}{12}} \psi_\hbar (U)\;,
 \\
 & \quad \Rightarrow \mathcal{Z}_b^{\U(1)_{-\frac{1}{2}}+\Phi} (m,\nu_0=0) \simeq  e^{\frac{m^2- 2\pi i (b+b^{-1})m}{4\pi i } + \frac{i \pi (b^2+b^{-2})}{12}} \mathcal{Z}_b^{\Phi} (m,\nu_0=0)\;.
 \end{split}
\end{align}
\paragraph{$k=-3/2$ Case :  SUSY Enhancement} The partition function is
\begin{align}
&\mathcal{Z}_{b} \big{(}m,\nu \big{)} = \int \!\frac{dZ}{\sqrt{2\pi \hbar}} \,e^{-\frac{Z^2+ 2Z\left(U +\nu \left(i \pi  + \frac{\hbar}2\right)\right)   }{2\hbar}} \psi_{\hbar} (Z)\bigg{|}_{U=bm}\; .
\end{align}
The saddle point equation and Bethe-vacua equation are 
\begin{align}
\begin{split}
&(\textrm{saddle point equation})\;:\;-Z^{(\alpha, \ell_z)}+\log (1-e^{-Z^{(\alpha, \ell_z)}}) -U+ 2\pi i \ell^\alpha_z - i\pi \nu=0\;,
\\
&(\textrm{Bethe-vacua equation})\;:\;\frac{1-\frac{1}{z}}z = (-1)^\nu e^U\;.
\end{split}
\end{align}
There are two Bethe-vacua $\{z^{\alpha}\}_{\alpha=1,2}$ and there are corresponding two saddle points $\{Z^{(\alpha, \ell_z)}(U,\nu)\}$ with a proper choice of $\ell_z \in \mathbb{Z}$. The round 3-sphere free energy $ F_{b=1}(m, \nu) := - \Re \log \mathcal{Z}_{b=1}(m,\nu)$ with general choice of the R-charge mixing $\nu$ can  be computed using eq.~\eqref{Z, k and CT} and \eqref{s0,s2 for u(1) +Phi}:
\begin{align}
\begin{split}
 &F_{b=1}(m, \nu)  =- \Re  \log \sum_{\alpha=1,2}\exp \bigg{(}s^{\alpha}_0 \big{(}m+2\pi i (\nu+1), \nu_0 = -1\big{)} \bigg{)}
 \\
 &=- \Re  \sum_{\alpha=1,2} \frac{2 \exp \left(\frac{i \text{Li}_2\left(e^{-Z}\right)}{2 \pi } - \frac{i Z^2}{4 \pi }+\frac{1 }{2} Z +(1-\ell_z^\alpha) Z  \right) }{-3 (e^Z-1)+e^Z+1}\bigg{|}_{ Z = Z^{(\alpha, \ell_z)}(U=m+2\pi i (\nu+1), \nu_0=-1)}\;.
 \end{split}
\end{align}
The free energy has $\mathbb{Z}_2$ symmetry
\begin{align}
F_{b=1}(m=0,\nu) = F_{b=1}(m=0, -\nu)\;,
\end{align}
and has maximum at 
\begin{align}
\nu_{\rm IR}=0\;.
\end{align}
At the IR fixed point $\nu_{\rm IR}=0$, there are two Bethe-vacua 
\begin{align}
\begin{split}
&z^{(\alpha=1)} = \frac{1}2 (-1+\sqrt{5})\;, \quad z^{(\alpha =2)} = \frac{1}2 (-1-\sqrt{5})\;.
\end{split}
\end{align}
Here we choose $\nu_0 = -1$ and $U = 2\pi i (\nu_{\rm IR}-\nu_0) = 2\pi i $. Basic CFT data at the IR fixed point can be computed using eq.~\eqref{Z, k and CT},\eqref{s0,s2 for u(1) +Phi} and \eqref{partial s0 for u(1) +Phi} with $k=-3/2, \nu_{0}=-1$ and $\nu_{\rm IR}=0$
\begin{align}
\begin{split}
&F = F_{b=1}(m=0, \nu =0) = - \Re \log  \left(\frac{1+ \exp ( \frac{3\pi i}5 )}{\sqrt{5}} \right) = 0.642965\;,
\\
&C_{JJ} = 8 \, \partial^2_m F_{b=1}(m=0, \nu =0) = \frac{2}{25} \left(8-\frac{5 \sqrt{2 \sqrt{5}+5}}{\pi }\right) = 0.248137 \;,
\\
&C_T = \frac{8}{\pi^2}  \partial^2_b F_{b=1}(m=0, \nu =0 )=  \frac{8}{25} \left(8-\frac{5 \sqrt{2 \sqrt{5}+5}}{\pi }\right) = 0.992549\;.
\end{split}
\end{align}
Note the equality $4C_{JJ}=C_T$ which is a strong evidence for the IR $\mathcal{N}=4$ supersymmetry of the theory. Refer to \cite{Gang:2018huc} for more evidences for the IR  enhancement. 

\subsection{\texorpdfstring{$\U(1)_k $ Coupled to Two  Chiral Multiplets of Charge $+1$}{U(1)k Coupled to Two Chiral Multiplets of Charge +1}}  The squashed 3-sphere partition function and the superconformal index are
\begin{align}
	\begin{split}
		&\mathcal{Z}_{b} \big{(}m,\nu \big{)} = \int \! \frac{dZ}{\sqrt{2\pi \hbar}} \, e^{\frac{(k+1) Z^2- 2 Z(U+ (i \pi  + \frac{\hbar}2) \nu )   }{2\hbar}} \psi_{\hbar} (Z) \psi_{\hbar}(Z)\bigg{|}_{U= bm }\; ,
		\\
		&\CI_q ({\bf m},u;\nu) = \sum_{{\bf n} \in \mathbb{Z}} \oint_{|v|=1} \frac{dv}{2\pi i v}\, v^{-\bf m} ((-q^{\frac{1}{2}})^{\nu}u )^{-\bf n} v^{(k+1){\bf n}} \CI_{\Delta}({\bf n},v;q)  \CI_{\Delta}({\bf n},v;q)\;.
	\end{split}
\end{align}
The theory has $\U(1)_J \times \SU(2)_\Phi$ flavor symmetry. $\U(1)_J$ is the topological symmetry while $\SU(2)_\Phi$ is the symmetry rotating the two chiral multiplets. The R-symmetry can be mixed only with the $U(1)_J$.  $m$ and $\nu$ are the real mass and the R-symmetry mixing parameter for the $U(1)_J$ respectively, while $({\bf m},u)$ are (background magnetic flux, fugacity) for the  $U(1)_J$. In the index formula, we turned off the (background magnetic flux, fugacity) for the $\SU(2)_\Phi$ symmetry. 
The proper CS level quantization is
\begin{align}
	k \in \mathbb{Z} \;.
\end{align}
The condition \eqref{criterion 2} is satisfied when 
\begin{align}
	\nu_0 + (k+1) \in 2\mathbb{Z}\;. 
\end{align}
The twisted superpotential at leading order is
\begin{align}
\CW_0^{\vec{\ell}} (Z,U,\nu)= \frac{(k+1)}2  Z^2+ 2 \textrm{Li}_2 (e^{-Z})+2 \pi i \ell_z Z- Z U - i \pi \nu Z\;. 
\end{align}
The saddle points and Bethe-vacua of the theory are determined by following  equations
\begin{align}
\begin{split}
&\textrm{Saddle point equation : }(k+1) Z + 2\log (1-e^{-Z}) - U - i \pi \nu + 2\pi i \ell_z =0\;,
\\
&\textrm{Bethe-vacua equation : } z^{k+1} \left(1-\frac{1}z\right)^2 = e^{U}(-1)^\nu\;.
\end{split}
\end{align}
For each Bethe-vacuum $z^{(\alpha)}$, there is an associated saddle point $Z^{(\alpha, \ell_z)}$ upon a proper choice of $\ell_z$. Up to 3-loop, the perturbative invariants are
\begin{align}
\begin{split}
&S_0(U,\nu) = \mathcal{W}_0^{\vec{\ell}}(Z,U,\nu) \;,
\\
&e^{2S_1} = \frac{z^{1-\nu}}{B (1-z)} \;,
\\
&S_2 (U,\nu=0)= \frac{  4 B^3-12 B^2 (z+1)+6 B (z^2+5 z)-20 z^2}{24 B^3 (z-1)}\;,
\\
&S_2 (U,\nu=-1) = \frac{4 B^3-3 B^2 \left(z^2-2 z+9\right)-6 B (z^2-7z)-20 z^2}{24 B^3 (z-1)}\;,
\\
&S_3 (U,\nu=0) = \frac{-2 B^5 z (z+3)+B^4 z \left(3 z^2+34 z+27\right)-B^3 z \left(z^3+49 z^2+127 z+27\right)}{24 B^6 (z-1)^2}
\\
&\qquad \qquad \qquad \; +\frac{2 B^2 z^2 \left(11 z^2+98 z+67\right)-20 B \left(5 z^4+11 z^3\right)+120 z^4}{24 B^6 (z-1)^2}\;,
\\
&S_3 (U,\nu=-1) = \frac{-8 B^5 z+16 B^4 z (z+3)-2 B^3 \left(5 z^3+65 z^2+32z\right)}{24 B^6 (z-1)^2}
\\
&\qquad \qquad \qquad\;\;+\frac{2 B^2 z^2 \left(z^2+62 z+113\right)-40 B \left(z^4+7 z^3\right)+120 z^4}{24 B^6 (z-1)^2}\;.
\end{split}
\end{align}
Here we define
\begin{align}
B := (k+1)z+1-k\;.
\end{align}
The $b\rightarrow 1$  expansion coefficients $s_n^{\alpha}(m,\nu)$ in eq.~\eqref{s0,s1,s2,s3}  are 
\begin{align}
\begin{split}
&\exp \big{(}s_0 (m,\nu) \big{)} =\frac{\exp \left(\frac{Z \left(i m-2 i \log \left(1-e^{-Z}\right)+\pi  (-3 \nu -2\ell_z+8)\right)+4 i \text{Li}_2\left(e^{-Z}\right)}{4 \pi }\right)}{B (1-z)}\;,
\\
& \partial_m s_0 (m,\nu) =\frac{i B m (z-1)+2 \pi  B (\nu +\nu  (-z)-2)+4 \pi  z}{2 \pi  B^2}\;,
\\
&\partial^2_m s_0 (m,\nu)=  \frac{i B^3 (z-1)+4 \pi  B^2 z+2 i B m (z-1) z-4 \pi  B z (\nu  (z-1)+z+3)+16 \pi  z^2}{2 \pi  B^4}\;, \label{sn for u(1) +2 Phi}
\end{split}
\end{align}
and
\begin{align}
\begin{split}
&s_2 (m,\nu=0) = \frac{1}{6 \pi  B^6 (z-1)^2} \bigg{(} 8 \pi^2 i B^6 \left(z-1\right) -960 \pi ^3 z^4
\\
&+B^5 \left(-8 i \pi ^2 m z^2-6 \pi  m z^2+8 i \pi ^2 m z+12 \pi  m z-6 \pi  m+16 \pi ^3 z^2-24 i \pi ^2 z^2+48 \pi ^3 z+24 i \pi ^2\right)
\\
&+B^4 \big{(}6 \pi  m^2 z^3-12 \pi  m^2 z^2+6 \pi  m^2 z+24 i \pi ^2 m z^3+6 \pi  m z^3+48 i \pi ^2 m z^2-12 \pi  m z^2-72 i \pi ^2 m z
\\
&\qquad \; +6 \pi  m z-24 \pi ^3 z^3+12 i \pi ^2 z^3-272 \pi ^3 z^2+48 i \pi ^2 z^2-216 \pi ^3 z-60 i \pi ^2 z \big{)}
\\
&+B^3 \big{(}i m^3 z^4-3 i m^3 z^3+3 i m^3 z^2-i m^3 z-6 \pi  m^2 z^4-6 \pi  m^2 z^3+30 \pi  m^2 z^2-18 \pi  m^2 z
\\
&\qquad  \;  -12 i \pi ^2 m z^4 -180 i \pi ^2 m z^3+84 i \pi ^2 m z^2+108 i \pi ^2 m z+8 \pi ^3 z^4+392 \pi ^3 z^3
\\
&\qquad \; -40 i \pi ^2 z^3+1016 \pi ^3 z^2+40 i \pi ^2 z^2+216 \pi ^3 z\big{)}
\\
&+B^2 \big{(}24 \pi  m^2 z^4-48 \pi  m^2 z^3+24 \pi  m^2 z^2+128 i \pi ^2 m z^4+192 i \pi ^2 m z^3-320 i \pi ^2 m z^2
\\
&\qquad \; -176 \pi ^3 z^4-1568 \pi ^3 z^3-1072 \pi ^3 z^2\big{)}
\\
&+B \left(-240 i \pi ^2 m z^4+240 i \pi ^2 m z^3+800 \pi ^3 z^4+1760 \pi ^3 z^3\right) \bigg{)} \;,
\\
&s_2 (m,\nu=-1) = \frac{1}{6 \pi  B^6 (z-1)^2} \bigg{(} 8 \pi^2 i B^6 \left( z-1\right) -960  \pi ^3 z^4
\\
&+B^5 \big{(}3 \pi  m z^3-8 i \pi ^2 m z^2-15 \pi  m z^2+8 i \pi ^2 m z+21 \pi  m z-9 \pi  m-6 i \pi ^2 z^3
\\
&\qquad \;  +18 i \pi ^2 z^2+64 \pi ^3 z-66 i \pi ^2 z+54 i \pi ^2\big{)}
\\
&+B^4 \big{(}6 \pi  m^2 z^3-12 \pi  m^2 z^2+6 \pi  m^2 z+6 \pi  m z^3+96 i \pi ^2 m z^2-12 \pi  m z^2 -96 i \pi ^2 m z+6 \pi  m z
\\
&\qquad \; -12 i \pi ^2 z^3-128 \pi ^3 z^2+96 i \pi ^2 z^2-384 \pi ^3 z-84 i \pi ^2 z\big{)}
\\
&+B^3 \big{(} i m^3 z^4-3 i m^3 z^3+3 i m^3 z^2-i m^3 z-24 \pi  m^2 z^3+48 \pi  m^2 z^2-24 \pi  m^2 z -120 i \pi ^2 m z^3
\\
&\qquad \;-72 i \pi ^2 m z^2 +192 i \pi ^2 m z+80 \pi ^3 z^3-40 i \pi ^2 z^3+1040 \pi ^3 z^2+40 i \pi ^2 z^2+512 \pi ^3 z \big{)}
\\
&+B^2 \big{(} 24 \pi  m^2 z^4-48 \pi  m^2 z^3+24 \pi  m^2 z^2+32 i \pi ^2 m z^4+384 i \pi ^2 m z^3
\\
&\qquad \; -416 i \pi ^2 m z^2-16 \pi ^3 z^4-992 \pi ^3 z^3-1808 \pi ^3 z^2 \big{)}
\\
&+B \left(-240 i \pi ^2 m z^4+240 i \pi ^2 m z^3+320 \pi ^3 z^4+2240 \pi ^3 z^3\right) \bigg{)}\;. \label{sn for u(1) +2 Phi 2}
\end{split} 
\end{align}
\paragraph{$k=0$ Case : $\SU(3)$ Symmetry Enhancement} In the case, the  $\U(1)_J \times \SU(2)_\Phi$ symmetry is enhanced to $\SU(3)$ at IR \cite{Gang:2017lsr,Gang:2018wek,Gaiotto:2018yjh,Benini:2018bhk}. Under the symmetry enhancement, the Cartan $\mathfrak{u}(1)_J$ for the topological $\U(1)_J$ symmetry is embedded into $\mathfrak{su}(3)$ as follows
\begin{align}
\mathfrak{u}(1)_J  = \frac{1}3\textrm{diag} \{1,1,-2 \} \in \mathfrak{su}(3)\;.
\end{align}
 Using eq.~\eqref{Z, k and CT},\eqref{sn for u(1) +2 Phi} and \eqref{sn for u(1) +2 Phi 2}, we can compute $F_{b=1}(m,\nu)$ and check that
\begin{align}
F_{b=1}(m=0,\nu) = F_{b=1}(m=0,2-\nu) \textrm{ is maximized at $\nu = \nu_{\rm IR} = 1$}.
\end{align}
At the IR fixed point $\nu_{\rm IR}=1$, there are two Bethe-vacua 
\begin{align}
	\begin{split}
		&z^{(\alpha=1)} =e^{i \pi/3}\;, \quad z^{(\alpha =2)} = e^{-i \pi/3}\;.
	\end{split}
\end{align}
Here, we choose $\nu_0 = 1$ and $U = 2\pi i (\nu_{\rm IR}-\nu_0) = 0$. 
Using eq.~\eqref{Z, k and CT},\eqref{sn for u(1) +2 Phi} and \eqref{sn for u(1) +2 Phi 2}, we can compute the basic CFT data of the SCFT at the fixed point
\begin{align}
\begin{split}
&F =F_{b=1}(m=0, \nu =1 )= -\log  \frac{1}{\sqrt{3}} \left( e^{\frac{V}{2\pi}} - e^{-\frac{V}{2\pi}}\right) \simeq 0.968723\;,
\\
& C_{JJ} = 8 \partial^2_m F_{b=1}(m=0, \nu =1)= \frac{4  \coth \left( \frac{V}{2\pi}\right)}{\sqrt{3}\pi} - \frac{16}9 \simeq 0.576242\;,
\\
&C_T = \frac{8}{\pi^2}  \partial^2_b F_{b=1}(m=0, \nu =1 ) = \frac{16}{27} \bigg{(} \frac{11 \sqrt{3} \coth \left( \frac{V}{2\pi}\right)}\pi - 6\bigg{)} \simeq 2.02706\;.
\end{split}
\end{align}
Here
\begin{align}
V := 2\, \textrm{Im}[\textrm{Li}_2 (e^{i \pi/3})] = 2.02988\;.
\end{align}
The  $F$ matches the  computation in \cite{Garoufalidis:2014ifa}. Note that the $C_T$ is less than that of the free theory of 3 chiral multiplets, which also has $\SU(3)$ symmetry.  

\section*{Acknowledgements}
We benefited tremendously from discussion with Victor Mikhaylov, whom we thank for initial collaboration
as well as many crucial inputs. The research of D.G.\ was supported in part by the National Research Foundation of Korea (NRF) grant 2019R1A2C2004880. The research of M.Y.\ was supported in part by WPI Research Center Initiative, MEXT, Japan, and by the JSPS Grant-in-Aid for Scientific Research (No.\ 17KK0087, No.\ 19K03820 and No.\ 19H00689).

\appendix
\section{Quantum Dilogarithm} \label{App : QDL}
The quantum dilogarithm function  (Q.D.L)   $\psi_\hbar (Z)$ is defined by  \cite{Faddeev:1993rs} ($\hbar = 2\pi i b^2$)
\begin{align}
\psi_\hbar (Z) := \begin{cases} \displaystyle\prod_{r=1}^\infty \frac{1-q^r e^{-Z}}{1-\tilde{q}^{-r+1}e^{-\tilde{Z}}}&\mbox{if } |q|<1\;, \\ 
\displaystyle \prod_{r=1}^\infty \frac{1-\tilde{q}^r e^{-\tilde{Z}}}{1-q^{-r+1}e^{-Z}} &\mbox{if } |q|>1 \;,\\ 
\end{cases} \label{def : Q.D.L}
\end{align}
with
\begin{align}
q:=e^{2\pi i b^2}\;, \quad \tilde{q}:=e^{2\pi i b^{-2}}\;, \quad \tilde{Z}:= \frac{1}{b^2}Z\;.
\end{align}
The function satisfies the following difference equations
\begin{align}
\psi_\hbar (Z + 2\pi i b^2) = (1-e^{-Z}) \psi_{\hbar} (Z)\;, \quad \psi_\hbar (Z + 2\pi i ) = \left(1-e^{-\frac{Z}{b^2}}\right) \psi_{\hbar} (Z)\;. \label{Difference eqns for QDL}
\end{align}
At the special value $b=1$, the Q.D.L simplifies as
\begin{align}
\log \psi_{\hbar=2 \pi i }(Z)  = \frac{-(2\pi +i Z) \log (1-e^{-Z})+i \textrm{Li}_2 (e^{-Z})}{2\pi} \;. \label{QDL at b=1}
\end{align}
Poles and zeros of Q.D.L are 
\begin{align}
\begin{split}
&\textrm{simple poles : } 2\pi i \mathbb{Z}_{\leq  0}+ 2\pi  i b^2 \mathbb{Z}_{\leq 0}\;,
\\
&\textrm{simple zeros : } 2\pi i  \mathbb{Z}_{\geq  1} + 2\pi  i b^2 \mathbb{Z}_{\geq 1}\;. \label{QDL : poles/zeros}
\end{split}
\end{align}
We have an integral representation:
\begin{align} 
\log\psi_\hbar(Z)=\int_{\mathbb{R}+i 0^+}  \frac{e^{ \frac{ i t Z }{\pi b} + t (b+b^{-1})}}{\sinh(b t)\sinh(b^{-1}t)} \frac{dt}{4t}\;, \quad \textrm{for $0<\textrm{Im}[Z]<2\pi (1+b^{2})$\;.} \label{QDL : integral}
\end{align}
The asymptotic  expansion when $\hbar =2\pi i b^2 \rightarrow 0$ is given by
\begin{align}
\log\psi_{\hbar}(Z) \xrightarrow{\;\;b^2\rightarrow 0\;\;}   \sum_{n=0}^{\infty} \frac{B_n \hbar^{n-1}}{n!} \textrm{Li}_{2-n}(e^{-Z}, \ell_0,\ell_z)\;. \label{asymptotic of QDL}
\end{align}
Here $B_n$ is the $n$-th Bernoulli number with $B_1=1/2$.

\newpage
\bibliographystyle{JHEP}
\bibliography{ref}

\end{document}